\documentclass[amsmath, amsfonts, showpacs, twocolumn, prb]{revtex4}
\usepackage{graphicx}
\usepackage{epsfig}
\usepackage{bm}
\usepackage{euscript}
\usepackage{dcolumn}
\usepackage{color}
\usepackage{amsmath}
\usepackage{amssymb}

\newcommand{\rt}{(\mathbf{r},t)}
\newcommand{\rtp}{(\mathbf{r},t')}
\newcommand{\rtpp}{(\mathbf{r},t'')}
\newcommand{\A}{\mathbf{A}}
\newcommand{\J}{\mathbf{j}}
\newcommand{\Eq}[1]{Eq.~\eqref{#1}}
\newcommand{\eq}[1]{\eqref{#1}}
\newcommand{\Div}{\,\mbox{div}}

\begin{document}

\title{Keldysh Ginzburg-Landau action of fluctuating superconductors}
\author{Alex Levchenko}
\author{Alex Kamenev}

\affiliation{Department of Physics, University of Minnesota,
Minneapolis, MN 55455, USA}

\date{June 19, 2007}

\begin{abstract}
We derive  Ginzburg-Landau action by systematically integrating out
electronic degrees of freedom in the framework of the Keldysh
nonlinear $\sigma$-model of disordered superconductors. The
resulting Ginzburg-Landau functional contains a nonlocal
$\Delta$-dependent contribution to the diffusion constant, which
leads, for example, to Maki-Thompson corrections. It also exhibits
an anomalous Gor'kov-Eliashberg coupling between $\Delta$ and the
scalar potential, as well as a peculiar nonlocal nonlinear term. The
action is gauge invariant and satisfies the fluctuation dissipation
theorem. It may be employed, e.g., for calculation of higher moments
of the current fluctuations.
\end{abstract}

 \maketitle

\section{Introduction}
\label{section-intro}

Time dependent Ginzburg-Landau (TDGL) theory has received a lot of
attention and was a subject of controversy over many years.
\cite{Schmid,Abrahams,Caroli-Maki,Gorkov-Eliashberg, Woo-Abrahams,
Eliashberg,Houghton-Maki,Hu-Thompson,Cyrot,Kramer,
Schon,Hu,Watts-Tobin,Krempasky,Otterlo} Gor'kov and
Eliashberg~\cite{Gorkov-Eliashberg} (GE) were probably among the
first who realized that the thermodynamic Ginzburg-Landau equation
may be generalized for the time-dependent phenomena in the case of
gapless superconductivity (see also earlier
publications~\cite{Schmid,Caroli-Maki}). The latter occurs either in
the presence of magnetic impurities, or in the fluctuating regime at
$T>T_c$. Notably  GE equation contained an anomalous nonlocal
coupling between the order parameter $\Delta$ and the scalar
potential: the fact that was frequently overlooked in many
subsequent treatments.

Extension of the TDGL theory to a gapped phase turns out to be a
very demanding problem. As noted by Gor'kov and Eliashberg, the
difficulty stems from the singularity of the BCS density of states
at the gap edge. The latter leads to a slowly decaying oscillatory
response at frequency $2\Delta/\hbar$ in the time domain.  As a
result, the expansion in powers of the small parameter $\Delta/T_c$
fails. In principle, it may be augmented by an expansion in
$\Delta/(\hbar\omega)$, in case the external fields are
high-frequency ones. To describe low-frequency responses in the
gapped phase, one needs a time nonlocal version of the TDGL theory.
The analysis is greatly simplified in the presence  of a
pair-breaking mechanism, such as magnetic impurities or energy
relaxation. Such a mechanism may eliminate singularity in the
density of states, leading to gapless phase in the presence of
finite $\Delta$. Under these conditions, an expansion in powers of
$\Delta\tau_{\phi}/\hbar$ and $\omega\tau_{\phi}$ is justified  and
thus a {\em time-local} TDGL equation may be derived (here
$\tau_{\phi}$ is the pair-breaking time). In the present work, we
restrict ourselves to the fluctuating regime $T>T_c$, where the
spectrum is gapless automatically and there is no need in an
explicit  pair-breaking mechanism.

Soon after the GE work, Aslamazov-Larkin (AL)
\cite{Aslamazov-Larkin} and Maki-Thompson (MT) \cite{Maki,Thompson}
corrections to conductivity of fluctuating superconductors were
discovered in the diagrammatic linear response framework. While  AL
term had naturally followed from TDGL theory (see, e.g.,
books~\cite{Abrikosov,Tinkham,Larkin-Varlamov}),  MT phenomena were
seemingly absent in  TDGL formalism.  Based on the
work,~\cite{Houghton-Maki} it was proposed~\cite{Cyrot} that in
order to include  MT term into the set of TDGL equations, one has to
substitute the renormalized  conductivity
$\sigma\rightarrow\sigma+\sigma^{MT}$ in the expression for the
current, supplementing TDGL equation. While leading to the correct
{\em static average} current (by construction), this way of handling
the problem fails to satisfy the fluctuation-dissipation theorem
(FDT). Indeed, it does not provide any prescription for calculating
{\em higher moments} of the current (even in equilibrium). Another
drawback of the approach of Ref.~[\onlinecite{Cyrot}] is that it
fails to incorporate a peculiar frequency dependence of  MT
phenomena, stemming from the time nonlocality of  MT terms. The
procedure introduced phenomenologically in Ref.~[\onlinecite{Cyrot}]
was latter elegantly derived in Ref.~[\onlinecite{Volkov-Nagaev}]
using nonequilibrium Green functions technique. Let us also mention
few other works where a combined set of TDGL and kinetic equations
was suggested.\cite{Krempasky,Hu} An imaginary-time action of
fluctuating superconductors was discussed in
Ref.~[\onlinecite{Otterlo}].

In the present publication, we derive a set of coupled {\em
stochastic} TDGL and Maxwell equations, which are suitable for
calculation of both average current and its higher moments. This set
of equations is an immediate consequence of the effective Keldysh
action written in terms of the fluctuating order parameter and
electromagnetic potentials. Technically, we employ the nonlinear
$\sigma$-model in the Keldysh representation \cite{KA,FLS} to
perform disorder averaging. We then systematically integrate out the
electronic degrees of freedom, neglecting Anderson localization
effects. The resulting {\em effective} action, written in terms of
the order parameter and electromagnetic potentials, naturally and
unmistakably contains both  MT terms and anomalous GE coupling
between the order parameter and electric field.

We restrict ourselves with the fluctuating regime $T>T_c$ only,
leaving the case $T<T_c$ (and magnetic impurities) for future
studies. As  always, the  Ginzburg-Landau treatment requires the
condition
\begin{equation}
T-T_c \ll T_c\, ,
                                                  \label{parameter}
\end{equation}
which is central to our consideration. We also assume that both the
order parameter and the electromagnetic fields  vary on the spatial
scale which is much larger than $\xi_0=\sqrt{D/T_c}$ (here $D$ is
the diffusion constant) and the time scale which is much slower than
$1/T_c$ (hereafter we adopt units, where $\hbar=c=1$). Moreover, we
shall rely on the fact that the electronic system is always in a
local thermal equilibrium. This in turn implies that the external
fields are not too large. More precisely, the electric field $\bf E$
is such that $eE\xi_0 \ll T_c$, while the magnetic field $\bf H$ is
restricted by the condition $eH\xi_0\ll 1/\xi_0$. Notice that these
conditions do {\em not} restrict our treatment to the linear
response regime. Nonlinear phenomena may be included, as long as a
characteristic scale of nonlinear effects satisfies the inequalities
given above.

The restrictions on  spatial and temporal scales of the external
fields  along with the fact that electrons are in local equilibrium
considerably simplify the theory. In particular, most of the terms
in the effective action acquire a local form in space and time.
Nevertheless, the effective theory does {\em not} take a completely
local form. The diffusion constant obtains a $\Delta$-dependent
contribution, with essentially nonlocal coupling to the order
parameter. If averaged over the fluctuations of the order parameter,
this nonlocal term yields  MT correction to conductivity. We note,
however, that  an average current is not the only manifestation of
the nonlocal term. The latter  also contributes to the current noise
as well as to its higher moments. Another nonlocal effect in the
effective action is the way the order parameter interacts with the
time-dependent electric field. This is the anomalous GE term. There
is one special gauge ($\mathcal{K}$-gauge), where an anomalous term
takes an especially simple form. In what follows, we shall explain
the $\mathcal{K}$-gauge and perform all the calculations in it. The
resulting action may be then transformed back into an arbitrary
gauge.

The use of the Keldysh formalism is important in several respects.
First, it allows to augment the replica trick to perform the
quenched disorder averaging procedure. Second and more important, it
is the only consistent way to derive real-time dynamics. The use of
the imaginary-time formalism, although possible, requires performing
the analytical continuation procedure. The latter is known to be
exceedingly demanding for MT as well as time nonlocal nonlinear
terms. Working directly in real time allows to make all the
expressions physically transparent, unobscured by the peculiarities
of the analytical continuation. Finally, the Keldysh formalism
naturally allows to extend the treatment to the situations, where
the assumption of local equilibrium is not applicable. Although not
considered in the present work, a treatment of a nonequilibrium
fluctuating superconductivity is a subject of great interest.

The rest of the paper is organized as follows. In the next section,
we present our main results in the form of the set of coupled
stochastic equations for the order parameter and electromagnetic
potentials. In  Sec. \ref{Section-Formalism} we introduce the basic
elements of the Keldysh nonlinear $\sigma$-model and explain the way
the effective action is derived by integrating out diffuson and
Cooperon degrees of freedom. Technical details of this procedure are
delegated to a number of appendixes. Finally, in Sec.
\ref{Section-Conclusions}, we summarize our findings and briefly
discuss their possible applications.

%------------------------------------------------------------------------------
%------------------------------------------------------------------------------

\section{Set of stochastic equations}
\label{Section-Results}

The most compact way to present our results is in the form of the
effective Keldysh action which is a functional of the fluctuating
order parameter and electromagnetic potentials. Since it requires
introducing some notations, we postpone discussion of the action
until Sec. \ref{Section-Formalism}. Here, we present an equivalent
way to display the same information using the set of {\em
stochastic} TDGL and Maxwell equations.

In  presence of the scalar  $\Phi\rt$ and  vector  $\A\rt$
potentials the complex order parameter $\Delta\rt$ obeys the
following TDGL equation:
\begin{equation}
 \big( \partial_{t}-2ie\, \partial_t \mathcal{K} \big) \Delta = \left[
D\left(\nabla-2ie\mathbf{A} \right)^{2} - \tau^{-1}_{GL}\right]
\Delta  +\xi_\Delta\, ,
                                          \label{TDGL}
\end{equation}
here $D$ is the diffusion constant, $e$ is the electron charge, and
\begin{equation}
\tau_{GL}=\frac{\pi}{8(T-T_c)}\,
                                          \label{tauGL}
\end{equation}
is the Ginzburg--Landau relaxation time. The field
$\mathcal{K}(\mathbf{r},t)$ satisfy the following equation:
\begin{equation}
\left(\partial_t - D\nabla^2 \right)\mathcal{K}\rt=\Phi\rt - D\Div
\mathbf{A}\rt\,.
                                          \label{Kdef}
\end{equation}
The complex Gaussian noise $\xi_\Delta(\mathbf{r},t)$ has the
correlator
\begin{equation}
\big\langle \xi_\Delta(\mathbf{r},t)\,
\xi^{*}_\Delta(\mathbf{r'},t') \big\rangle = {16\, T^{\,2}\over
\pi\nu}\,\, \delta_{\mathbf{r}-\mathbf{r'}}\, \delta_{t-t'}\, ,
                                           \label{xiDelta}
\end{equation}
where $\nu$ is the density of states. Unlike   TDGL equation
frequently found in the literature,~\cite{Houghton-Maki,Hu-Thompson,
Abrikosov,Tinkham,Larkin-Varlamov} the lhs. of Eq.~(\ref{TDGL})
contains GE anomalous term \cite{Gorkov-Eliashberg} $\partial_t
\mathcal{K}\rt$ instead of the scalar potential $\Phi\rt$. The two
coincide in the limit of spatially uniform potentials, cf.
Eq.~(\ref{Kdef}). In a generic case, they are rather distinct and
$\mathcal{K}\rt$ is a nonlocal functional of the scalar and the
longitudinal vector potentials. The standard motivation behind
writing the scalar potential $\Phi\rt$ in the lhs of TDGL is the
gauge invariance. Notice, however, that a local gauge transformation
\begin{eqnarray}
\Delta&\to& \Delta\,e^{-2ie\chi}\,;\quad\quad
\Phi\to\Phi-\partial_t\chi\,;\nonumber \\
 \A&\to&\A-\nabla\chi\,; \quad\,\,\,\,\,\,  \mathcal{K}\to \mathcal{K}-\chi
                                                  \label{gauge}
 \end{eqnarray}
leaves Eq.~\eqref{TDGL} unchanged and therefore this form of TDGL
equation is perfectly gauge invariant. The last expression in Eq.
\eq{gauge} is an immediate consequence of \Eq{Kdef} and the rules of
transformation for $\Phi\rt$ and $\A\rt$.

We have suppressed the nonlinear terms in Eq.~(\ref{TDGL}), since
they are of lesser importance for $T>T_c$. A   detailed discussion
of the nonlinear terms is presented in section Sec.
\ref{section-nonlinear}. We note, however, that in addition to the
conventional $|\Delta|^2\Delta$ local term, there is other
essentially nonlocal and time-dependent nonlinear term in TDGL
equation.

TDGL equation (\ref{TDGL}) takes an especially simple form in the
{\em $\mathcal{K}$-gauge}, which is obtained by choosing
$\chi\rt=\mathcal{K}\rt$ in Eq.~(\ref{gauge}). In other words, the
gauge is specified by the relation
\begin{equation}
\Phi_\mathcal{K}-D\Div \A_\mathcal{K}=0\, ,
                                                   \label{Kgauge}
\end{equation}
where $\Phi_\mathcal{K}=\Phi-\partial_t \mathcal{K}$ and
$\A_\mathcal{K}=\A-\nabla \mathcal{K}$. In such a gauge, the
anomalous term in the lhs of TDGL is absent and the latter obtains
the form
\begin{equation}
\partial_{t} \Delta_\mathcal{K} = \left[
D\left(\nabla-2ie\mathbf{A_\mathcal{K}} \right)^{2} -
\tau^{-1}_{GL}\right] \Delta_\mathcal{K} + \xi_\Delta\, ,
                                          \label{KTDGL}
\end{equation}
where $\Delta_\mathcal{K}=\Delta \, e^{-2ie\mathcal{K}}$. Employing
Eq.~(\ref{Kgauge}) along with  the expression for the electric field
$\mathbf{E}=\partial_t \A_\mathcal{K}-\nabla \Phi_\mathcal{K}$, one
finds for the vector potential $\A_\mathcal{K}\rt$  in the rhs of
TDGL equation (\ref{KTDGL}),
\begin{equation}
\A_\mathcal{K}\rt = \A_{\perp}\rt + \A_{\parallel\mathcal{K}}\rt
\, .
                                                      \label{A_K}
\end{equation}
Here, $\A_{\perp}$ is the gauge invariant {\em transverse} part of
the vector potential and the {\em longitudinal}  part in the
$\mathcal{K}$-gauge is given by
\begin{equation}
\A_{\parallel\mathcal{K}}\rt =  \int\! \mathrm{d} \mathbf{r}'
\mathrm{d} t'\, \mathcal{D}^{\,\mathbf{r},\mathbf{r}'}_{t,t'}
\mathbf{E}_{\parallel}(\mathbf{r}',t')\, ,
                                                      \label{A_Kparrallel}
\end{equation}
where $\mathbf{E}_{\parallel}$ is the  longitudinal part of the
electric field. The kernel
$\mathcal{D}^{\,\mathbf{r},\mathbf{r}'}_{t,t'}\sim\theta(t-t')$ is
the retarded  Green function of the diffusion operator
\begin{equation}
(\partial_{t}-D\nabla^{2})\mathcal{D}^{\mathbf{r},\mathbf{r}'}_{t,t'}
=\delta_{t-t'} \delta_{\mathbf{r}-\mathbf{r}'}.
                                                     \label{D_diff}
\end{equation}

In addition to the equation for the order parameter, the complete
theory must provide two material equations for the current $\J\rt$
and charge  $\rho\rt$ densities. The first of these equations is the
continuity relation:
\begin{equation}
\Div\, \J+\partial_t \rho =0\,.
                                                      \label{cont}
\end{equation}
As for the second one, we found the following expression for the
current density:
\begin{widetext}
\begin{equation}
\J\rt=\int\mathrm{d}t' \left[D\delta_{t-t'} +\delta
D^{MT}_{\mathbf{r},t,t'} \right]\left[  e^2\nu
\mathbf{E}(\mathbf{r},t') - \nabla \rho(\mathbf{r},t')\right] +
\frac{\pi e\nu D}{4T}\ {\rm
Im}\left[\Delta^*_\mathcal{K}\rt(\nabla-2ie\A_\mathcal{K})
\Delta_\mathcal{K}\rt\right]+\mathbf{\xi}_{\J}\rt\, .
                                                    \label{current}
\end{equation}
The nonlocal part of the diffusion coefficient is the functional of
the order parameter (as well as the electromagnetic potentials)  and
is given by the expression
\begin{equation}
\delta D^{MT}_{\mathbf{r},t,t'}[\Delta_\mathcal{K}] = \frac{\pi
D}{4T}\int\mathrm{d} \mathbf{r}'\mathrm{d}\mathbf{r}''\
 \mathcal{C}^{\mathbf{r},\mathbf{r}'}_{\tau,t,t'}
 {\Delta}^*_\mathcal{K}\left(\mathbf{r}',\tau\right)
 {\Delta}_\mathcal{K}(\mathbf{r}'',\tau)\
\bar{\mathcal{C}}^{\mathbf{r}'',\mathbf{r}}_{\tau,t',t}\ ,
                                                 \label{deltaD}
\end{equation}
with $\tau=(t+ t')/2\,$.  The retarded
$\mathcal{C}^{\,\mathbf{r},\mathbf{r}'}_{\tau,t,t'}\sim
\theta(t-t')$  and advanced
$\bar{\mathcal{C}}^{\,\mathbf{r},\mathbf{r}'}_{\tau,t,t'}\sim
\theta(t'-t)$  Cooperon propagators are Green functions of the
following equations:
\begin{subequations}\label{cooperon}
\begin{equation}
\left[\phantom{-}\partial_t
-ie\Phi_{\mathcal{K}}(\mathbf{r},\tau_+)+
ie\Phi_{\mathcal{K}}(\mathbf{r},\tau_-)-D\left[\nabla-
ie\A_\mathcal{K}(\mathbf{r},\tau_+)-
ie\mathbf{A}_\mathcal{K}(\mathbf{r},\tau_-)\right]^2 \right]
\mathcal{C}^{\mathbf{r},\mathbf{r}'}_{\tau,t,t'} =
\delta_{\mathbf{r}-\mathbf{r}'}\delta_{t-t'}\ ,
\end{equation}
\begin{equation}
\left[-\partial_t+ie\Phi_{\mathcal{K}}(\mathbf{r},\tau_+)-
ie\Phi_{\mathcal{K}}(\mathbf{r},\tau_-)-D\left[\nabla-
ie\A_\mathcal{K}(\mathbf{r},\tau_+)-
ie\mathbf{A}_\mathcal{K}(\mathbf{r},\tau_-)\right]^2 \right]
\bar{\mathcal{C}}^{\mathbf{r},\mathbf{r}'}_{\tau,t,t'} =
\delta_{\mathbf{r}-\mathbf{r}'}\delta_{t-t'}\,,
\end{equation}
\end{subequations}
where $\tau_\pm=\tau\pm t/2$. Note that the MT term obeys
causality, since $\delta D^{MT}_{\mathbf{r},t,t'}\sim
\theta(t-t')$, and  gauge invariant in view of \Eq{gauge}. Being
averaged  over the fluctuations of the order parameter
$\langle\delta D^{MT}\rangle_\Delta$, it leads to the
(frequency-dependent) Maki-Thompson correction to the
conductivity. Equation~(\ref{current}) is more general, however,
as it allows to calculate the higher moments of the current as
well. The current fluctuations are induced by the stochastic term
in the TDGL equation $\xi_\Delta$ as well as by the current noise
$\xi_{\mathbf{j}}\rt$ given by the Gaussian vector process with
the correlator:
\begin{equation}
\big\langle \xi_\J^\alpha(\mathbf{r},t)\,
\xi_\J^\beta(\mathbf{r'},t') \big\rangle =\delta_{\alpha,\beta}\,
T e^2 \nu \left(2D \delta_{t-t'}+\delta D^{MT}_{\mathbf{r},t,t'} +
\delta D^{MT}_{\mathbf{r},t',t}\right)
\delta_{\mathbf{r}-\mathbf{r'}}\, ,
                                           \label{xij}
\end{equation}
guaranteeing validity of FDT, here $\alpha,\beta=x,y,z$. Equations
\eqref{KTDGL}, \eqref{cont}, and \eqref{current} must be
supplemented by  Maxwell equations for the electromagnetic fields.
In the next section, we show how these results may be derived from
the microscopic model.
\end{widetext}
%-----------------------------------------------------------------------------------
%-----------------------------------------------------------------------------------

\section{Keldysh sigma model formalism}
\label{Section-Formalism}

\subsection{Notations and the $\sigma$-model action}

We employ   Keldysh technique, \cite{Keldysh} which allows to go
beyond the linear response  and is formulated directly in real time.
The formalism considers the evolution along the closed contour in
the time direction. It thus deals with the two ``replica'' of each
field, one encoding the evolution in the forward and another in the
backward time direction. It is convenient to introduce half-sum and
half-difference of these fields to which we shall refer as {\em
classical} and {\em quantum} components correspondingly.
\cite{Kamenev} As a result, all the fields acquire the vector
structure, e.g., the scalar potential
$\vec{\Phi}=(\Phi^{cl},\Phi^{q})$, the vector potential
$\vec{\mathbf{A}}=(\mathbf{A}^{cl},\mathbf{A}^{q})$ and the complex
order parameter  $\vec{\Delta}=(\Delta^{cl},\Delta^{q})$. It is also
convenient to introduce  $4\times 4$ matrix notations for these
fields in the space which is a direct product of Keldysh and Nambu
spaces
\begin{subequations}
\begin{equation}\label{A-Phi}
\check{\Phi}=[\Phi^{cl}\sigma_{0}+
\Phi^{q}\sigma_{x}]\otimes\tau_{0},\
\check{\mathbf{A}}=[\mathbf{A}^{cl}\sigma_{0}+
\mathbf{A}^{q}\sigma_{x}]\otimes\tau_{0},
\end{equation}
\begin{equation}\label{Delta}
\check{\Delta}=[\Delta^{cl}\sigma_{0}+\Delta^{q}\sigma_{x}]\otimes\tau_{+}-
[\Delta^{*cl}\sigma_{0}+\Delta^{*q}\sigma_{x}]\otimes\tau_{-},
\end{equation}
\end{subequations}
here $\sigma_\alpha$ and $\tau_\alpha$ are sets of Pauli matrices
in  Keldysh and Nambu spaces correspondingly ($\alpha=0,x,y,z$)
and $\tau_{\pm}=(\tau_{x}\pm i\tau_{y})/2$.

Our starting point is the nonlinear $\sigma$-model, \cite{KA,FLS}
which systematically takes care of the elastic disorder averaging.
In the framework of this formalism, the electron dynamics is
described by the field $\check{Q}(\mathbf{r};t,t')$ which is a
matrix in the $4\times 4$ Keldysh-Nambu space as well as an infinite
matrix (integral kernel) with respect to its two time indices. For a
short-range correlated disorder (the only case considered here) the
$Q$ matrix is a local function of the spacial variable $\mathbf{r}$.
The $Q$ matrix obeys the local nonlinear constraint
\begin{equation}
\check{Q}^{\, 2}(\mathbf{r})=1\, ,
                                                 \label{Qsquare}
\end{equation}
where $\check{Q}^{\, 2}$ is understood as the matrix multiplication
in $4\times 4$ as well as in the time space and the rhs is the unit
operator in this space.

It is very convenient~\cite{KA,NAA,FLS} to single out the gauge
degree of freedom of the $Q$-matrix field by expressing  it in the
following form:
\begin{equation}\label{Gauge-Q-K}
\check{Q}(\mathbf{r};t,t')=
e^{ie\check{\Xi}\check{\mathcal{K}}(\mathbf{r},t)}\,
\check{Q}_{\mathcal{K}}(\mathbf{r};t,t')\,\,
e^{-ie\check{\mathcal{K}}(\mathbf{r},t')\check{\Xi}}\, .
\end{equation}
Here  $\check{\mathcal{K}}(\mathbf{r},t)$ is a scalar gauge field in
the matrix representation analogous to Eq.~(\ref{A-Phi}),
$\check{\Xi}=\sigma_{0}\otimes\tau_{z}$, and
$\check{Q}_{\mathcal{K}}$ is the new $Q$-matrix field free from the
gauge ambiguity. Obviously, the field $\check{Q}_{\mathcal{K}}$ also
satisfy  the nonlinear constraint $\check{Q}^{2}_{\mathcal{K}}=1$.
In what follows, we shall use the freedom of choosing the gauge
field $\mathcal{K}$ to adjust a saddle point on the $Q$-manifold,
Eq.~(\ref{Qsquare}), according to local scalar and vector
potentials. Therefore, the $\vec{\mathcal{K}}$ field should be
understood  as a certain functional of the electromagnetic
potentials $\vec{\Phi}$ and $\vec{\mathbf{A}}$ which fixes a special
gauge.

The  Keldysh nonlinear $\sigma$-model, we employ here,  was
formulated for normal metals  by Kamenev and Andreev~\cite {KA} and
extended for superconductors by Feigelman \textit{et al}.~\cite{FLS}
Its action takes the following form:
\begin{subequations}\label{S}
\begin{equation}
S[Q,\Delta,\mathbf{A},\Phi]=S_{\Delta}+S_{\Phi}+S_{\sigma},
\end{equation}
\begin{equation}\label{S-Delta-Phi}
S_{\Delta}=-\frac{\nu}{2\lambda}\,
\mathrm{Tr}\left[\check{\Delta}_{\mathcal{K}}
\check{\Upsilon}\check{\Delta}_{\mathcal{K}}\right],\quad
S_{\Phi}=\frac{e^{2}\nu}{2}\, \mathrm{Tr}
\left[\check{\Phi}_{\mathcal{K}}\check{\Upsilon}
\check{\Phi}_{\mathcal{K}}\right],
\end{equation}
\begin{equation}\label{S-sigma}
S_{\sigma}\! =\frac{i\pi\nu}{8}\,
\mathrm{Tr}\left[D(\partial_{\mathbf{r}}
\check{Q}_{\mathcal{K}})^{2} \! - \!
\check{\Xi}\partial_{t}\check{Q}_{\mathcal{K}}\! + \!
4ie\check{\Phi}_{\mathcal{K}}\check{Q}_{\mathcal{K}} \! + \!
4i\check{\Delta}_{\mathcal{K}}\check{Q}_{\mathcal{K}}\right],
\end{equation}
\end{subequations}
here $\lambda$ is  superconductive coupling constant,
$\check{\Upsilon}=\sigma_{x}\otimes\tau_{0}$, and the covariant
spatial derivative is defined according to
\begin{equation}\label{Covariant-Derivative}
\partial_{\mathbf{r}}\check{Q}_{\mathcal{K}}
=\nabla_{\mathbf{r}}\check{Q}_{\mathcal{K}}-
ie[\check{\Xi}\check{\mathbf{A}}_{\mathcal{K}},\check{Q}_{\mathcal{K}}]\,.
\end{equation}
The subscript $\mathcal{K}$ denotes gauge transformed fields
\begin{subequations}
\begin{equation}\label{Gauge-A-Phi-K}
\check{\Phi}_{\mathcal{K}}=\check{\Phi}-\partial_{t}\check{\mathcal{K}},
\quad \check{\mathbf{A}}_{\mathcal{K}}=
\check{\mathbf{A}}-\nabla\check{\mathcal{K}},
\end{equation}
\begin{equation}
\hskip 1.5cm \check{\Delta}_{\mathcal{K}}(\mathbf{r},t)=
e^{-ie\check{\Xi}\check{\mathcal{K}}(\mathbf{r},t)}
\check{\Delta}(\mathbf{r},t)\,
e^{ie\check{\mathcal{K}}(\mathbf{r},t)\check{\Xi}}\, .
\end{equation}
\end{subequations}
The trace operation in Eq. (\ref{S}) $\mathrm{Tr}[\ldots]$ implies
integration over the space and time indices as well as matrix trace
in the $4\times 4$ Keldysh-Nambu space. The action written above
should be supplemented by the standard Maxwell term
$S_M=\mathrm{Tr}\left[ \check{\mathbf E} \check{\Upsilon}
\check{\mathbf E}+\check{\mathbf H} \check{\Upsilon} \check{\mathbf
H}\right]/16\pi$.

Our eventual goal is to integrate out fluctuations of the electronic
degrees of freedom represented by the field $\check Q_{\mathcal{K}}$
to end up with an effective action in terms of the electromagnetic
potentials and the order parameter only. To this end, one needs a
parametrization of the $\check Q_{\mathcal{K}}$ field which
explicitly resolves the nonlinear constraint (\ref{Qsquare}).
Following Refs.~[\onlinecite{KA,FLS}], we adopt the exponential
parametrization
\begin{equation}\label{Exp-Parameterization}
\check{Q}_{\mathcal{K}}(\mathbf{r}) = e^{-
\check{W}(\mathbf{r})/2}\, \check{\Lambda}\,\,
e^{\check{W}(\mathbf{r})/2 }\,,
\end{equation}
where the matrix multiplication in the time space is implicitly
assumed. The matrix $\check{\Lambda}$ represents the normal metal
saddle point (hereafter we work at $T>T_c$) in the absence of
external fields,
\begin{equation}\label{Lambda}
\check{\Lambda}_{t,t'}= \left(\begin{array}{cc}\delta_{t-t'}&
2F_{t-t'}\\
0&-\delta_{t-t'}\end{array}\right)\otimes\tau_z
=\check{\mathcal{U}}\check{\Lambda}_0
\check{\mathcal{U}}^{-1},
\end{equation}
where $\check{\Lambda}_0=\sigma_z\otimes\tau_z$ and
\begin{equation}\label{U}
\check{\mathcal{U}}_{t,t'}=\check{\mathcal{U}}_{t,t'}^{-1}=
\left(\begin{array}{cc}\delta_{t-t'}&
F_{t-t'}\\
0&-\delta_{t-t'}\end{array}\right)\otimes\tau_0\ .
\end{equation}
The function $F_{t-t'}$ is the Fourier transform of the equilibrium
distribution function ${F}_{\varepsilon}=\tanh(\varepsilon/2T)$,
i.e.,
\begin{equation}
F_t= -\frac{iT}{\sinh(\pi T t)} \stackrel{t\gg
1/T}{\longrightarrow} {i\,\over 2T}\,\, \delta'(t)\, .
                                                      \label{Ftime}
\end{equation}
The last expression is an approximation applicable for slowly
varying external fields. Note that choosing the parametrization in
the form [Eq.~(\ref{Exp-Parameterization})] does {\em not} imply
that the electrons are in the state of the global thermal
equilibrium. Indeed, the actual distribution function is given by
$e^{ie{\mathcal{K}^{cl}}(\mathbf{r},t)} F_{t-t'}
e^{-ie{\mathcal{K}^{cl}}(\mathbf{r},t')}$, cf.
Eq.~(\ref{Gauge-Q-K}), and includes local variations (e.g., chemical
potential) due to the presence of electromagnetic potentials. The
field $\vec{\mathcal{K}}$ is to be chosen (see below) to achieve
this goal in an optimal way.

The matrix field $\check{W}_{t,t'}(\mathbf{r})$ in
Eq.~(\ref{Exp-Parameterization}) represents fluctuations of the
electronic degrees of freedom and is to be integrated out. To avoid
redundancy of the parametrization, one needs to ensure that the
matrix $\check{W}$ does not commute with $\check{\Lambda}$. This is
achieved by requiring that
$\check{W}\check{\Lambda}+\check{\Lambda}\check{W}=0$. This
condition is resolved by introducing four real fields
$w^{\alpha}_{tt'}(\mathbf{r}),\bar{w}^{\alpha}_{tt'}(\mathbf{r})$
with $\alpha=0,z$ representing diffuson degrees of freedom  and two
complex fields $w_{tt'}(\mathbf{r}),\bar{w}_{tt'}(\mathbf{r})$ for
Cooperon degrees freedom. (The bar symbol denotes an independent
field, {\em not} a complex conjugation.) These fields are built into
the matrix~\cite{FLS}
\begin{equation}\label{W}
\check{W}=\check{\mathcal{U}}\check{\mathcal{W}}
\check{\mathcal{U}}^{-1},\quad\,\,\,
\check{\mathcal{W}}=\left(\begin{array}{cc}
w\tau_{+} - w^{*}\tau_{-} & w_{0}\tau_{0}+w_{z}\tau_{z}\\
\bar{w}_{0}\tau_{0}+\bar{w}_{z}\tau_{z} & \bar{w}\tau_{+} -
\bar{w}^{*}\tau_{-}
\end{array}\right),
\end{equation}
where the asterisk stays for complex conjugation and the matrix
$\check{\mathcal{U}}$ is defined in Eq.~(\ref{U}).

%-----------------------------------------------------------------------------------
\subsection{Diffuson modes, $\mathcal{K}$--gauge, and normal action }
\label{section-diffuson}

In this subsection, we shall disregard the fluctuations of the order
parameter $\vec{\Delta}$. Since we are  not interested in the
weak-localization effects, we can disregard the Cooperon degrees of
freedom $w_{tt'}(\mathbf{r}),\bar{w}_{tt'}(\mathbf{r})$ in the
matrix (\ref{W}) as well. We then substitute the matrix $\check{W}$,
Eq.~(\ref{W}), written in terms of the diffuson fields
$w^{\alpha}_{tt'}(\mathbf{r}),\bar{w}^{\alpha}_{tt'}(\mathbf{r})$
($\alpha=0,z$) into the sigma-model action (\ref{S-sigma}) and
expand it to the {\em linear} order in the diffuson fields. We focus
first on the $z$-components $w^{z},\bar{w}^{z}$. Demanding that the
terms linear in $w^{z},\bar{w}^{z}$ vanish, one obtains the
condition~\cite{KA} (for details see Appendix
\ref{Appendix-Diffusons})
\begin{equation}\label{Vanishing-wz-condition}
[\check{\Phi}_{\mathcal{K}},\check{\Lambda}]+
D(\check{\Xi}\nabla\check{\mathbf{A}}_{\mathcal{K}}-
\check{\Lambda}\check{\Xi}\nabla\check{\mathbf{A}}_{\mathcal{K}}
\check{\Lambda})=0.
\end{equation}
This matrix equation may be resolved by a proper choice of the
gauge doublet
$\vec{\mathcal{K}}=(\mathcal{K}^{cl},\mathcal{K}^{q})$, thus
fixing the $\mathcal{K}$--gauge. Employing
Eqs.~(\ref{Gauge-A-Phi-K}) and (\ref{Lambda}), one may rewrite
Eq.~(\ref{Vanishing-wz-condition}) as an explicit gauge-fixing
condition
\begin{equation}
\vec{\Phi}_\mathcal{K}(\mathbf{r},\omega) =
\!\left(\begin{array}{cc}1 & -2B_{\omega} \\
0 & -1 \end{array}\right) D\,\mbox{div}
\vec{\mathbf{A}}_\mathcal{K}(\mathbf{r},\omega)\,,
                                        \label{Kgauge-complete}
\end{equation}
with
\begin{equation}\label{B}
B_{\omega}=\coth(\omega/2T) \stackrel{\omega \ll
T}{\longrightarrow} {2T/\omega}
\end{equation}
being the equilibrium bosonic distribution function. In the absence
of the quantum components of the fields (used to generate
observables), Eq.~(\ref{Kgauge-complete}) is reduced to the gauge
condition (\ref{Kgauge}) written for the classical field components.
It is important, however, to fix the gauge for both quantum and
classical components.

Equation (\ref{Kgauge-complete}) completes the task of finding the
gauge field and combined with Eqs. \eqref{Gauge-Q-K} and
\eqref{Lambda} provides the approximate saddle point, which is
determined for any given realization of the fields $\vec\Phi\rt$ and
$\vec\A\rt$. This general scheme guarantees that in the expansion
over $\check{W}$--fluctuations,  terms such as
$\mathrm{Tr}[\check{\Phi}_{\mathcal{K}}\check{W}]$ and
$\mathrm{Tr}[\check{\A}_{\mathcal{K}}\check{W}]$ do not appear in
the action.

This procedure does not completely eliminate terms linear in the
diffuson generators. Indeed, contributions of the form
$\mathrm{Tr}[\check{\Lambda}\check\A_{\mathcal{K}}
\check{\Lambda}\check{W}\check\A_{\mathcal{K}}]$ come from the
diamagnetic term of the $\sigma$-model action
(\ref{S-sigma}).~\cite{KA} Such terms are linear in the diffuson
fields $w^{0},\bar{w}^{0}$ and {\em quadratic} in the
electromagnetic potentials. Integrating out diffusons
$w^{0},\bar{w}^{0}$ in the Gaussian approximation yields a nonlocal
vertex {\em quartic} in the electromagnetic potentials. It is
exactly this quartic vertex which is responsible for
Altshuler-Aronov correction to the conductivity of  normal
metals~\cite{AA} (for details see Ref.~[\onlinecite{KA}]). Since
Altshuler-Aronov corrections do not exhibit a singular temperature
dependence in the vicinity of $T_{c}$, we shall ignore these terms
hereafter. It is an interesting and open question to investigate
other possible implications of these nonlinear terms.

Once the terms linear in diffuson generators are eliminated by a
choice of the proper  gauge, one may substitute the metallic saddle
point $\check{Q}_{\mathcal{K}}=\check{\Lambda}$ into the sigma-model
action [Eq.~(\ref{S})] to obtain the effective action in terms of
the electromagnetic potentials  (for details see Appendix
\ref{Appendix-Diffusons}). Such a procedure neglects nonlinear
interactions of the diffuson modes and thus amounts to disregarding
the Anderson localization effects. The resulting action takes the
form
\begin{equation}\label{S-N}
S_{N}= e^{2}\nu\ \mathrm{Tr}\left[\vec\Phi_{\mathcal{K}}^\dagger
\sigma_x \vec\Phi_{\mathcal{K}}+ \vec{\A}_{\mathcal{K}}^\dagger
\hat{\EuScript{T}}_D \vec{\A}_{\mathcal{K}}\right]\, ,
\end{equation}
where the operator $\hat{\EuScript{T}}_D$ is defined as
\begin{equation}\label{TD}
\hat{\EuScript{T}}_D=\left(\begin{array}{cc} 0 &
-\overleftarrow{\partial_{t}}D \\ -D\overrightarrow{\partial_{t}}
& 4iTD
\end{array}\right)\, .
\end{equation}
The arrows $\leftrightarrows$ on top of the time derivative imply
that  the differentiation is performed to the left/right,
respectively. Employing gauge fixing condition
[Eq.~(\ref{Kgauge-complete})], the action may be rewritten as
\begin{equation}\label{S-N-alternative}
\!\! \!\!\! S_{N}= e^{2}\nu D\,
\mathrm{Tr}\left[\vec{\A}_{\mathcal{K}}^\dagger
\left(\begin{array}{cc} 0 & D\nabla^2 -\overleftarrow{\partial_{t}} \\
D\nabla^2 - \overrightarrow{\partial_{t}} & 4iT
\end{array}\right) \vec{\A}_{\mathcal{K}} \right] .
\end{equation}

This action may now be employed to determine (fluctuating) charge
and current densities. To this end, one needs to introduce an
auxiliary vector Hubbard-Stratonovich field
$\xi_{\mathbf{j}}(\mathbf{r},t)$,  to decouple the term quadratic in
quantum component of the vector potential $\A_{\mathcal{K}}$
\begin{equation}
e^{-{4e^2\nu D T}\ {\rm Tr} \big[\A_{\mathcal{K}}^q\big]^2} =\int
\mathrm{D}[\xi_{\mathbf{j}}] \, e^{ -({4e^2\nu D T})^{-1} {\rm Tr}
\xi_{\mathbf{j}}^2 - 2i {\rm Tr}\A_{\mathcal{K}}^q
\xi_{\mathbf{j}}}.
                                              \label{HS}
\end{equation}
The resulting action is now linear in both $\Phi_{\mathcal{K}}^q$
and $\A_{\mathcal{K}}^q$ fields, allowing to define the charge and
current densities as
\begin{subequations}\label{rho-j}
\begin{equation}
\rho(\mathbf{r},t)=\frac{1}{2}\, \frac{\delta
S[\vec\Phi,\vec{\A}]}{\delta\Phi^{q}(\mathbf{r,t})}\ ,
\end{equation}
\begin{equation}
\mathbf{j}(\mathbf{r},t)=\frac{1}{2}\, \frac{\delta
S[\vec\Phi,\vec{\A}]}{\delta\mathbf{A}^{q}(\mathbf{r,t})}\, .
\end{equation}
\end{subequations}
It is important to note that the differentiation  here has to be
performed over the \textit{bare} electromagnetic potentials, while
the action is written in terms of the \textit{gauged} ones. The
connection between those $\{\vec\Phi,\vec\A \}\leftrightarrows
\{\vec\Phi_{\mathcal{K}},\vec\A_{\mathcal{K}}\}$ is provided by the
functional $\vec{\mathcal{K}}[\vec\Phi,\vec\A]$, which is implicit
in Eq.~(\ref{Kgauge-complete}) and in the explicit form is presented
in the Appendix \ref{Appendix-Diffusons}. A simple algebra then
leads to a set of the continuity equation (\ref{cont}) and the
expression for the normal current density
\begin{equation}\label{current-normal}
\mathbf{j}(\mathbf{r},t)=D\left[e^{2}\nu\mathbf{E}(\mathbf{r},t)-\nabla
\rho(\mathbf{r},t)\right]+\xi_{\mathbf{j}}(\mathbf{r},t)\, .
\end{equation}
The Hubbard-Stratonovich field $\xi_{\mathbf{j}}\rt$ has a meaning
of the Gaussian Langevin noise source~\cite{KS} with the correlation
function given by (cf. Eq.~(\ref{HS}))
\begin{equation}\label{Langevin}
\langle\xi^{\alpha}_{\mathbf{j}}(\mathbf{r},t)
\xi^{\beta}_{\mathbf{j}}(\mathbf{r}',t')\rangle=2 T e^{2}\nu D
\delta_{\alpha,\beta}\delta_{t-t'}
\delta_{\mathbf{r}-\mathbf{r}'}\, .
\end{equation}
Notice that because of assumed local equilibrium of  electronic
degrees of freedom, Eqs.~(\ref{current-normal}) and (\ref{Langevin})
do not lead to any excess noise beyond the one prescribed by
equilibrium FDT. This is not  the case for fluctuating
superconductors. Indeed,  the order parameter may be driven out of
equilibrium, while the electrons are still in the state of the local
equilibrium.

%------------------------------------------------------------------------------------
%------------------------------------------------------------------------------------
\subsection{Cooperon modes and superconducting  action}
\label{Section-Central}

Having taking care of the diffuson modes with the help of the
$\mathcal{K}$-gauge, we turn now to the fluctuations of the Cooperon
modes $w,\bar{w}$.  The latter are induced by the fluctuating order
parameter  $\vec\Delta\rt$.   To eliminate the Cooperon degrees of
freedom of the electronic system, we substitute parametrization
(\ref{W}) into the action (\ref{S-sigma}) and expand it to the
second order in $w,\bar{w}$. Once again, neglecting the higher order
terms in the expansion amounts to disregard the localization
effects. As a result, one obtains the following quadratic action
(for  details  see appendix~\ref{Appendix-Gradient-Expansion}):
\begin{widetext}
\begin{equation}\label{S-W}
S_{\sigma}[\check W,\vec{\Delta}_\mathcal{K},
\vec{\Phi}_{\mathcal{K}},\vec{\mathbf{A}}_{\mathcal{K}}] = i
\frac{\pi\nu}{4}\, \mathrm{Tr}\left[
\vec{\mathfrak{W}}^{\dag}_{tt'}(\mathbf{r})
\hat{\mathfrak{C}}^{-1}_{tt'}\vec{\mathfrak{W}}_{t't}(\mathbf{r})
-
2i\vec{V}^{\dag}_{tt'}(\mathbf{r})\vec{\mathfrak{W}}_{t't}(\mathbf{r})+
2i\vec{\mathfrak{W}}^{\dag}_{tt'}(\mathbf{r})\vec{V}_{t't}(\mathbf{r})
\right] ,
\end{equation}
here we have introduced vector
$\vec{\mathfrak{W}}_{tt'}(\mathbf{r})=[w_{tt'}(\mathbf{r}),
\bar{w}_{tt'}(\mathbf{r})]^{T}$, defined in the two dimensional
space of the complex Cooperon fields and the vector
$\vec{V}_{tt'}(\mathbf{r})=[V_{tt'}(\mathbf{r}),-\bar{V}_{tt'}(\mathbf{r})]^{T}$,
with the elements
\begin{equation}\label{V}
V_{tt'}(\mathbf{r})=\delta_{t-t'}\Delta^{cl}_{\mathcal{K}}\rt+
\frac{i}{2T}\delta'_{t-t'}\Delta^{q}_{\mathcal{K}}\rtp,\quad
\bar{V}_{tt'}(\mathbf{r})=\delta_{t-t'}\Delta^{cl}_{\mathcal{K}}\rt-
\frac{i}{2T}\delta'_{t-t'}\Delta^{q}_{\mathcal{K}}\rt.
\end{equation}
The factor $i\delta'(t-t')/(2T)$ multiplying the quantum component
of the order parameter is nothing but the long time approximation
for the fermionic distribution function $F_{t-t'}$,
Eq.~(\ref{Ftime}).  This approximation is adopted throughout the
subsequent calculations. The electromagnetic field--dependent
Cooperon matrix propagator has the following structure:
\begin{equation}\label{C}
\hat{\mathfrak{C}}^{-1}_{tt'}[\mathbf{A}_{\mathcal{K}}]=
\left[\begin{array}{cc} \mathcal{C}^{-1}_{tt'} & 0
\\ 0 &  \bar{\mathcal{C}}^{-1}_{tt'}
\end{array}\right] +
\left[\begin{array}{cc} \EuScript{N}_{tt'} & \EuScript{M}_{tt'}
\\ \EuScript{M}_{tt'} &  \EuScript{N}_{tt'}
\end{array}\right]
\end{equation}
with the matrix elements
\begin{subequations}\label{C-Matrix-Elements}
\begin{equation}
                                \label{C-1}
\mathcal{C}^{-1}_{tt'}=-\partial_{t}+\partial_{t'}+
ie[\Phi^{cl}_{\mathcal{K}}\rt-\Phi^{cl}_{\mathcal{K}}\rtp]-
D\left[\nabla-ie\mathbf{A}^{cl}_{\mathcal{K}}\rt-
ie\mathbf{A}^{cl}_{\mathcal{K}}\rtp\right]^{2},
\end{equation}
\begin{equation}
                                 \label{Cbar-1}
\bar{\mathcal{C}}^{-1}_{tt'}=\phantom{-}\partial_{t}-\partial_{t'}-
ie[\Phi^{cl}_{\mathcal{K}}\rt-\Phi^{cl}_{\mathcal{K}}\rtp]
-D\left[\nabla-ie\mathbf{A}_{\mathcal{K}}^{cl}\rt-
ie\mathbf{A}^{cl}_{\mathcal{K}}\rtp\right]^{2},
\end{equation}
\begin{equation}
                                    \label{N}
\EuScript{N}_{tt'}=-\delta_{t-t'}\frac{2eD}{T} \left[
\frac{1}{2}\Div\,\mathbf{A}^{q}_{\mathcal{K}}\rt +
\mathbf{A}^{q}_{\mathcal{K}}\rt\left[\nabla-
2ie\mathbf{A}^{cl}_{\mathcal{K}}\rt\right] \right],
\end{equation}
\begin{equation}
                                       \label{M}
\EuScript{M}_{tt'}= -2e^{2}D\left[\mathbf{A}^{q}_{\mathcal{K}}\rt+
\frac{i}{2T}\partial_{t}\mathbf{A}^{cl}_{\mathcal{K}}\rt\right]
\mathbf{A}^{q}_{\mathcal{K}}\rtp\, .
\end{equation}
\end{subequations}
\end{widetext}

We show below that the diagonal elements of
$\hat{\mathfrak{C}}^{-1}$ operator are responsible for the
conventional part of the TDGL theory in the form derived by
GE.~\cite{Gorkov-Eliashberg} More precisely, the terms
${\mathcal{C}}^{-1}$ and $\bar{\mathcal{C}}^{-1}$ yield  the TDGL
equation for the order parameter, while the additional diagonal
terms $\EuScript{N}$ (proportional to the quantum component of the
vector potential) lead to the superconductive part of the current.
Interestingly, this is not the entire story yet! Indeed, the
operator $\hat{\mathfrak{C}}^{-1}$ contain also the off-diagonal
elements $\EuScript{M}$, which induce cross correlations between $w$
and $\bar{w}$ Cooperon generators (i.e., they induce correlations
between rotations of retarded and advanced sectors of the $Q$-matrix
in the Keldysh space). It is very difficult (if not impossible) to
capture such terms within the analytical continuation technique of
GE. It is exactly these off-diagonal terms which are responsible for
MT contribution to the average current. We shall derive the
corresponding part of the effective TDGL action, which allows to
include MT effect in the higher moments of the current as well.

The next step is conceptually simple. It involves the Gaussian
integration over the Cooperon degrees of freedom (i.e., vector
$\vec{\mathfrak{W}}$) in Eq.~(\ref{S-W}). The result may be
schematically represented as $S\sim \mbox{Tr}[\vec V^\dagger
\hat{\mathfrak{C}}[\vec \A_{\mathcal{K}}]\, \vec V]$, which is an
action quadratic in the order parameter $\Delta$. One needs thus a
way to invert the operator $\hat{\mathfrak{C}}^{-1}$ given by
Eqs.~(\ref{C}) and (\ref{C-Matrix-Elements}). To this end, we notice
that the second term in the rhs of Eq.~(\ref{C}) contains the
quantum component of the vector potential and thus may be regarded
as small. Taking advantage of this fact, we first find the saddle
point of the action (\ref{S-W}) without the last term in the rhs of
Eq.~(\ref{C}) and then substitute this approximate saddle point into
the $\EuScript{N}$ and $\EuScript{M}$ terms. Taking variation with
respect to $w$ and $\bar{w}$, one finds for the approximate saddle
point
\begin{subequations}\label{w-saddle-point}
\begin{equation}
w_{\tau+{\eta\over 2},\tau-{\eta\over 2}}(\mathbf{r})={2\over i}\!
\int\!\! \mathrm{d} \mathbf{r}' \mathrm{d} \eta'\,
\mathcal{C}^{\mathbf{r},\mathbf{r}'}_{\tau,\eta,\eta'}
V_{\tau+{\eta'\over 2},\tau-{\eta'\over 2}}(\mathbf{r}')\, ,
\end{equation}
\begin{equation}
\bar{w}_{\tau+{\eta\over 2},\tau-{\eta\over 2}}(\mathbf{r})=2i\!
\int\!\! \mathrm{d} \mathbf{r}' \mathrm{d} \eta'\,
\bar{\mathcal{C}}^{\mathbf{r},\mathbf{r}'}_{\tau,\eta,\eta'}
\bar{V}_{\tau+{\eta'\over 2},\tau-{\eta'\over 2}}(\mathbf{r}')\,
\end{equation}
\end{subequations}
and  similar expressions for  conjugated Cooperon generators
$w^{*}$ and $\bar{w}^{*}$.  The retarded ($\eta>\eta'$) and
advanced ($\eta<\eta'$)  Cooperon propagators are determined by
Eqs.~(\ref{cooperon}).

\begin{figure}
\includegraphics[width=9cm]{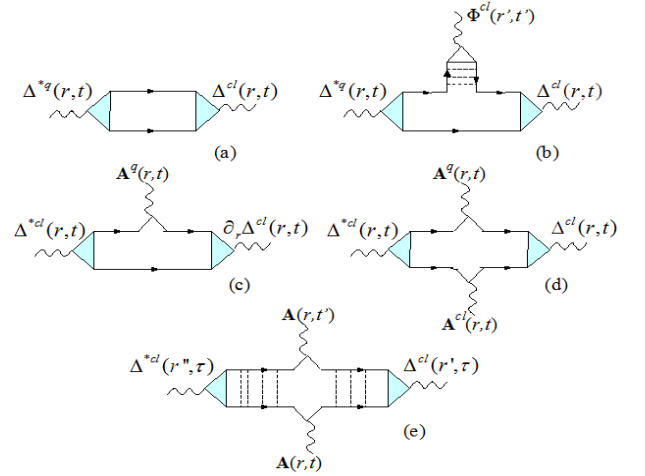}
\caption{(Color online) Diagrammatic representation of the effective
action, Eq.~\eqref{S-eff}. (a) Conventional Ginzbirg-Landau
functional $S_{GL}$. (b) Anomalous GE coupling between the scalar
potential  and the order parameter. (c) Gradient part of
supercurrent  action $S_{SC}$. (d) Diamagnetic component of the
supercurrent. (e) Nonlocal MT term (here $\tau=(t+t')/2$ and there
are two possibilities: one vector potential \textit{classical} and
another \textit{quantum}, which is part of the current, or two
\textit{quantum}, which is  FDT counterpart). \label{Fig1}}
\end{figure}

We now substitute the  saddle point Eq. \eqref{w-saddle-point} back
into the action \eqref{S-W}, to obtain the effective  action of a
fluctuating superconductor (details are presented in
Appendix~\ref{Appendix-Gradient-Expansion})
\begin{eqnarray}\label{S-eff}
\!\!\!\!\! \!\!\!\!\!
S_{\mathrm{eff}}[\vec\Delta_{\mathcal{K}},\vec\Phi_{\mathcal{K}},
\vec\A_{\mathcal{K}}] &=& S_{GL}[\vec\Delta_{\mathcal{K}},
\vec\A_{\mathcal{K}}]
+S_{SC}[\vec\Delta_{\mathcal{K}},\vec\A_{\mathcal{K}}] \nonumber \\
&+&
S_{MT}[\vec\Delta_{\mathcal{K}},\vec\A_{\mathcal{K}}]+S_{N}[\vec\Phi_{\mathcal{K}},
\vec\A_{\mathcal{K}}].
\end{eqnarray}
The last term here $S_{N}$ is the normal action \eqref{S-N}, or
equivalently (\ref{S-N-alternative}),  originating from the diffuson
degrees of freedom. The other three terms originate from the
Cooperon action (\ref{S-W}) in the way outlined above. Specifically,
the Ginzburg-Landau action $S_{GL}$ comes from the
${\mathcal{C}}^{-1}$, $\bar{\mathcal{C}}^{-1}$ terms in  the action,
the supercurrent action $S_{SC}$ from the $\EuScript{N}$ terms and
Maki-Thompson action $S_{MT}$ originates from the off-diagonal
$\EuScript{M}$ terms. The diagrammatic representation of these terms
is given in Fig.~\ref{Fig1}.

The TDGL part of the action
$S_{GL}[\vec\Delta_{\mathcal{K}},\A_{\mathcal{K}}^{cl}]$ has a local
form standard for the Keldysh formalism. It comes  from
$\mathrm{Tr}[\vec V^\dagger \mathcal C \vec V]$ and
$\mathrm{Tr}[\vec V^\dagger \bar{\mathcal C} \vec V]$ terms,
Fig.~\ref{Fig1}(a)
\begin{equation}\label{S-GL}
S_{GL} =  \frac{\pi\nu}{8T}\,\,  \mathrm{Tr}
\left[\vec{\Delta}^{\dag}_{\mathcal{K}}\rt \hat{L}^{-1}\,
\vec{\Delta}_{\mathcal{K}}\rt\right ]\ ,
\end{equation}
where the fluctuations propagator $\hat{L}$ has a typical bosonic
form in the Keldysh space
\begin{equation}\label{L-matrix}
\hat{L}^{-1}=\left(\begin{array}{cc}0 & L^{-1}_{A}
\\ L^{-1}_{R} & 4iT \end{array}\right) .
\end{equation}
Here, retarded and/or advanced components of the fluctuation
propagator are given by
\begin{equation}
L^{-1}_{R(A)}=\mp\partial_{t}- \tau^{-1}_{GL}+D
\left[\nabla-2ie\mathbf{A}^{cl}_{\mathcal{K}}\rt\right]^{2}\,,
                                                   \label{L}
\end{equation}
while Keldysh component of the propagator satisfy the FDT in
equilibrium
$L_{K}^{-1}=B_\omega\left(L^{-1}_{R}-L^{-1}_{A}\right)\to 4iT$ if
$\omega\ll T$. Note that the scalar potential
${\Phi}_{\mathcal{K}}^{cl}$, although present in the action
(\ref{S-W}) through the operators~(\ref{C-1}) and (\ref{Cbar-1}),
does not show up in the Ginzburg-Landau action (\ref{S-GL}). This
happens because upon substitution of the Cooperon generators by
their saddle point values (\ref{w-saddle-point}) the terms
$\Phi_{\mathcal{K}}^{cl}\rt$ and $\Phi_{\mathcal{K}}^{cl}\rtp$ in
Eqs.~(\ref{C-1}) and (\ref{Cbar-1}) cancel each other. [To be
precise there is a small residual term $\sim \partial_t
\Phi_{\mathcal{K}}^{cl}/T$, which we do not keep, since it exceeds
the accuracy of our calculations. For the same reason, terms with
$\Phi_{\mathcal{K}}^{q}$ are not kept in the $\EuScript{N}$ operator
Eq.~(\ref{N})].  As a result, the effective TDGL action depends only
on the vector potential, but {\em not} on the scalar potential, if
written in the $\mathcal{K}$-gauge. In any other gauge, there is a
linear coupling between the scalar potential and the $z$-diffuson
mode $\sim F\Phi^{cl}\bar{w}^z+\Phi^q w^z$, cf.
Eq.~(\ref{Vanishing-wz-condition}). Taken together with the terms
$\Delta \bar{w}w^z$ and $\Delta w \bar{w}^z$ (see next subsection)
and being averaged over the diffuson fluctuations, these terms lead
to a nonlocal coupling between the scalar potential $\Phi$ and the
order parameter $|\Delta|^2$, see Fig.~\ref{Fig1}(b). This is the
anomalous GE term. Thanks to the condition
(\ref{Vanishing-wz-condition}) the anomalous term is absent in the
$\mathcal{K}$-gauge, making this gauge especially convenient to work
in.

The supercurrent action
$S_{SC}[\vec\Delta_\mathcal{K},\vec\A_{\mathcal{K}}]$ comes from the
diagonal terms in Eq.~(\ref{S-W}) $\mathrm{Tr}[w\EuScript{N}w]$ and
$\mathrm{Tr}[\bar w\EuScript{N}\bar w]$, where Cooperon generators
$w$ and $\bar w$ are given by Eq.~(\ref{w-saddle-point}). It is also
local and given by
\begin{equation}\label{S-J}
S_{SC}=  \frac{\pi e\nu
D}{2T}\,\,\mathrm{Tr}\left\{\mathbf{A}^{q}_{\mathcal{K}}
\mathrm{Im}\left[\Delta^{*cl}_{\mathcal{K}}
(\nabla-2ie\mathbf{A}^{cl}_{\mathcal{K}})
\Delta^{cl}_{\mathcal{K}}\right]\right\}\,
.
\end{equation}
The gradient part of the supercurrent originates from
$\frac{1}{2}\Div\,\mathbf{A}^{q}_{\mathcal{K}} +
\mathbf{A}^{q}_{\mathcal{K}}\nabla$ terms in the rhs of
Eq.~(\ref{N}), while the diamagnetic current originates from $
2ie\mathbf{A}^{q}_{\mathcal{K}}\mathbf{A}^{cl}_{\mathcal{K}}$ term,
see Figa.~\ref{Fig1}(c) and \ref{Fig1}(d).

Time locality of the Ginzburg-Landau and the supercurrent actions
stems from their diagonal nature. Indeed, they both involve products
such as $w_{t,t'}w_{t',t}$. According to Eqs.~(\ref{V}) and
(\ref{w-saddle-point}), $w_{t,t'}\sim\tilde\theta(t-t')$, where
$\tilde\theta(t)$ is the step function smeared at the scale $1/T$.
Therefore, the time variables  in $w_{t,t'}w_{t',t}$ are compatible
only in the narrow vicinity of $t=t'$, i.e., $|t-t'|\lesssim 1/T$,
hence the time locality. This argument does {\em not} apply to the
off-diagonal MT term. Indeed, the latter involves product of
retarded and advanced generators $w_{t,t'}\bar{w}_{t',t}$ whose time
variables  are compatible for any $t>t'$.

The Maki-Thompson action
$S_{MT}[\Delta_\mathcal{K}^{cl},\vec\A_{\mathcal{K}}]$, coming from
the off-diagonal blocks of  $\hat{\mathfrak{C}}^{-1}$ operator, has
essentially time nonlocal form as explained above,
Fig.~\ref{Fig1}(e),
\begin{equation}\label{S-MT}
S_{MT} =  e^{2}\nu\,\mathrm{Tr}
\left[\vec{\mathbf{A}}^{\dag}_{\mathcal{K}} (\mathbf{r},t)
\hat{\EuScript{T}}_{\delta D}(t,t')
\vec{\mathbf{A}}_{\mathcal{K}}(\mathbf{r},t')] \right]\, ,
\end{equation}
where the operator $\hat{\EuScript{T}}_{\delta D}(t,t')$ is given by
[cf. Eq.~(\ref{TD})]
\begin{equation}\label{TdeltaD}
\hat{\EuScript{T}}_{\delta D}=\left(\begin{array}{cc} 0 &
-\overleftarrow{\partial_{t}}\,
\delta D^{MT}_{\mathbf{r},t',t} \\
-\delta D^{MT}_{\mathbf{r},t,t'}\, \overrightarrow{\partial_{t'}}
& 2iT \left( \delta D^{MT}_{\mathbf{r},t,t'} + \delta
D^{MT}_{\mathbf{r},t',t} \right)
\end{array}\right),
\end{equation}
where the $\delta D^{MT}[\Delta_\mathcal{K}^{cl}]$ functional is
given by Eq.~(\ref{deltaD}). Note that the MT action has {\em
exactly} the same structure as the second term in the normal action
(\ref{S-N}). It therefore amounts to the time nonlocal
renormalization of the {\em normal}  diffuson constant
$D\delta_{t-t'}\to D\delta_{t-t'}+\delta D^{MT}_{\mathbf{r},t,t'}$.

Finally, we comment on the so called density of states (DOS)
contributions.~\cite{Larkin-Varlamov} They originate from the
subleading terms (in characteristic frequency over temperature) in
the diagonal operator $\EuScript{N}$ [not written explicitly in
Eq.~(\ref{N})], see Appendix \ref{Appendix-Dos-contribution} for
details. Accounting for them leads to a {\em local} renormalization
of the density of states prefactor in the {\em normal} action,
Eq.~(\ref{S-N}), or Eq.~(\ref{S-N-alternative}),
\begin{equation}\label{deltaD-DOS}
\nu \rightarrow \nu_{\mathbf{r},t}= \nu \left[1-
\frac{7\zeta(3)}{4\pi^{2}T^{2}}\,
|\Delta^{cl}_{\mathcal{K}}\rt|^{2}\right],
\end{equation}
where $\zeta(x)$ is the Riemann zeta function.  This is  a small
effect in the regime we are working in.

%-----------------------------------------------------------------------------
%-----------------------------------------------------------------------------

\subsection{Nonlinear terms in the Ginzburg-Landau functional}
\label{section-nonlinear}

The nonlinear in $\Delta$ terms of the TDGL equation are not very
significant at $T>T_c$. Nevertheless, we shall discuss them here for
completeness. There are several ways ``$|\Delta|^4$'' terms appear
in the effective TDGL action. We shall keep track of
$\Delta^{*q}\Delta^{cl}|\Delta^{cl}|^2$ terms which directly
contribute to TDGL equation for $\Delta^{cl}$, discussing other
combinations only briefly. The most important way such terms appear
is through the {\em third} order expansion of the
$\mathrm{Tr}[\check{\Delta}_{\mathcal{K}}\check{Q}_{\mathcal{K}}]$
term of the $\sigma$-model action \eqref{S-sigma} in powers of
$\check{W}$. Keeping only the Cooperon generators and employing
$\mathrm{Tr}[\tau_{\pm}\tau_{z}\tau_{\mp}\tau_{\pm}\tau_{\mp}]=\mp
1$, one obtains
\begin{equation}\label{S-NL-trace}
S_{NL}=\frac{\pi\nu}{12}\,
\mathrm{Tr}\left[\Delta^{cl}_{\mathcal{K}}\left( w^{*}ww^{*}-
\bar{w}^{*}\bar{w}\bar{w}^{*}\right) +c.c.\right].
\end{equation}
Similar terms coming with $\Delta^{q}_{\mathcal{K}}$ component of
the order parameter eventually cancel out between $w$ and $\bar{w}$
contributions and thus are omitted in Eq.~(\ref{S-NL-trace}). Next,
one substitutes the saddle point value of the Cooperon generators
Eq.~\eqref{w-saddle-point}  into the action \eqref{S-NL-trace}, and
perform traces over the time indices. In doing so, one should keep
only the first power of the quantum component of the order parameter
[coming from Eq.~\eqref{V}]. The diagrammatic representation of the
corresponding terms is shown in Fig.~\ref{Fig2}(a). After
straightforward algebra (see Appendix \ref{Appendix-S-NL} for
details), one finds
\begin{equation}\label{S-NL-conventional}
S_{NL} = -\frac{7\zeta(3)\nu}{8\pi^{2}T^{2}}\, \mathrm{Tr}\left[
\Delta^{*q}_{\mathcal{K}}\rt\Delta^{cl}_{\mathcal{K}}\rt
|\Delta^{cl}_{\mathcal{K}}\rt|^{2}+ c.c.\right].
\end{equation}
This term is to be added to the retarded and advanced (but not
Keldysh) parts of the Ginzburg-Landau action (\ref{S-GL}). It leads
to the standard nonlinear term of the Ginzburg-Landau equation,
\cite{Abrikosov,Larkin-Varlamov} which is (a) local, (b) disorder
independent, in agreement with Anderson theorem.

\begin{figure}
\includegraphics[width=9cm]{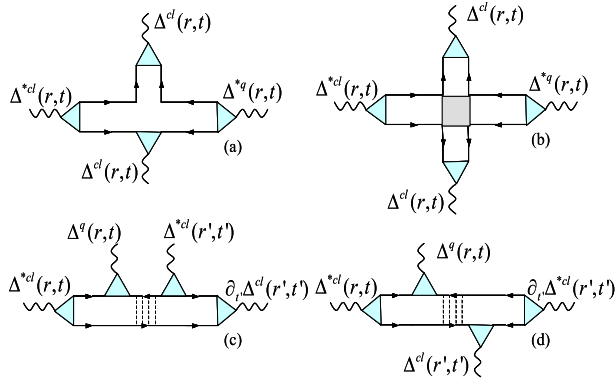}
\caption{(Color online) Nonlinear contributions to  Ginzburg-Landau
functional. (a) Diagram for the local nonlinear term of the action
Eq.\eqref{S-NL-conventional}. (b) Renormalization of the diffusion
constant in the TDGL equation -- shaded square is the Hikami box.
[(c) and (d)] terms involving diffuson channel, leading to a
nonlocal renormalization of the dynamic part in TDGL equation.
\label{Fig2}}
\end{figure}

Interestingly, this is not the only way the nonlinear terms appear
in the effective action. Let us mention two other venues. (i) One
may expand the $D\ \mathrm{Tr}[\nabla\check{Q}_{\mathcal{K}}]^2$
term of the nonlinear $\sigma$ model action (\ref{S-sigma}) to the
{\em fourth} order in the Cooperon fields, generating the so-called
Hikami box, see Fig.~\ref{Fig2}(b). One then substitutes the saddle
point value of the Cooperon generators \eqref{w-saddle-point} in
such a term to obtain a  contribution quartic in the order
parameter. This term leads to a  local renormalization of the
diffusion constant in TDGL equation $\delta D_{\mathbf{r},t}\sim
D|\Delta^{cl}\rt|^2/T^2$. There is {\em no} MT nonlocal
renormalization of the diffusion constant in the {\em
superconductive} part of the action (as opposed to the {\em normal
one} where both MT and DOS renormalizations take place). (ii) There
is yet another source of nonlinear terms (we are not aware if it had
been discussed previously in the literature). It originates as a
result of mixing between Cooperon and diffuson channels.  To see it,
one expands
$\mathrm{Tr}[\check{\Delta}_{\mathcal{K}}\check{Q}_{\mathcal{K}}]$
term of the $\sigma$-model action (\ref{S-sigma})  to the {\em
second}  order in $\check{W}$. This way one generates interaction
vertices  of the following structure
$\mathrm{Tr}[\check{\Delta}_{\mathcal{K}} w^* ({w}^{\alpha}\pm
\bar{w}^\alpha)]+ c.c.$ with $\alpha=0,z$ and corresponding terms
with $\bar{w}$. We then perform Gaussian integration over the
diffuson fields $w^{\alpha},\bar{w}^\alpha$ to obtain a nonlocal
vertex $\mathrm{Tr}[w^*_{tt'}
\check{\Delta}_{\mathcal{K}}(\mathbf{r},t') \langle w^\alpha_{t't}
\bar{w}^\alpha_{\tilde t \tilde t'} \rangle
\check{\Delta}_{\mathcal{K}}^*(\mathbf{r}',\tilde t') w_{\tilde t'
\tilde t}]$, where $\langle w^\alpha \bar{w}^\alpha \rangle$ is the
diffuson propagator, Eq.~(\ref{D_diff}). There is a similar vertex
with $\bar{w}^*,\bar{w}$ generators. Such a nonlocal vertex is
effectively a renormalization of the {\em diagonal} part of the
$\hat{\mathfrak{C}}^{-1}_{tt'}$ operator in Eq.~(\ref{S-W}). It is
important to stress that the diffuson admixture does {\em not}
generate the off-diagonal terms in $\hat{\mathfrak{C}}^{-1}_{tt'}$
operator, thus not affecting directly the MT channel. We then
substitute the saddle point values of the Cooperon generators,
Eq.~(\ref{w-saddle-point}), in this nonlocal vertex and find
\begin{widetext}
\begin{equation}
\widetilde{S}_{NL}= \frac{7\zeta(3)\nu}{8\pi^{2}T^{2}}\mathrm{Tr}\
[\Delta^{*q}_{\mathcal{K}}\rt \Delta^{cl}_{\mathcal{K}}\rt
\mathcal{D}^{\mathbf{r},\mathbf{r}'}_{t,t'}\Delta^{*cl}_{\mathcal{K}}
(\mathbf{r}',t')
\, \partial_{t'}\Delta^{cl}_{\mathcal{K}}(\mathbf{r}',t')
+c.c.]\,.
                                          \label{S-NL-new}
\end{equation}
\end{widetext}
This term is formally of the same order of magnitude as the
conventional one, Eq.~(\ref{S-NL-conventional}). Indeed, the
diffuson propagator $\mathcal{D}^{\mathbf{r},\mathbf{r}'}_{t,t'}$ is
not cut by the temperature [unlike the Cooperon propagators in
Fig.~\ref{Fig2}b].

%----------------------------------------------------------------------------------
%----------------------------------------------------------------------------------

\subsection{Equations of motion and Coulomb interactions}
\label{section-Coulomb}

To derive the stochastic equations of motion presented in Sec.
\ref{Section-Results}, one needs to get rid of terms quadratic in
quantum components of the fields: $\Delta_\mathcal{K}^q$ in $S_{GL}$
and $\A_\mathcal{K}^q$ in $S_N+S_{MT}$. This is achieved with the
Hubbard-Stratonovich transformation similar to Eq.~(\ref{HS}) for
$\A_\mathcal{K}^q$ and
\begin{equation}\label{HS-transformation}
e^{-{\pi\nu\over 2}
\mathrm{Tr}|\Delta^{q}_{\mathcal{K}}|^{2}}=\int\mathrm{D}
[\xi_{\Delta}]e^{-\frac{\pi\nu}{8T}\mathrm{Tr}\left[\frac{|\xi_{\Delta}|^{2}}{4T}
-i\xi^{*}_{\Delta}\Delta^{q}_{\mathcal{K}}-i\xi_{\Delta}
\Delta^{*q}_{\mathcal{K}}\right]}
\end{equation}
for $\Delta_\mathcal{K}^q$. As a result, the effective action
(\ref{S-eff}) acquires the form linear in quantum components of the
fields. Integration over the  latter leads to the functional delta
functions imposing the stochastic equations of motion. This way the
TDGL equation (\ref{KTDGL}), which we present here including the
nonlinear terms
\begin{widetext}
\begin{equation}\label{KTDGL-NL}
\left[\partial_{t} +\tau^{-1}_{GL} -
D(\nabla-2ie\mathbf{A}_{\mathcal{K}}\rt)^{2}+ {b\over T}
|\Delta_{\mathcal{K}}\rt|^{2} - {b\over T}
\!\int\!\mathrm{d}t'\mathrm{d}\mathbf{r}'
\Delta^{*}_{\mathcal{K}}(\mathbf{r}',t')
\partial_{t'}\Delta_{\mathcal{K}}(\mathbf{r}',t')\,
\mathcal{D}^{\mathbf{r},\mathbf{r}'}_{t,t'}
\right]\Delta_{\mathcal{K}}\rt=\xi_{\Delta}\rt
\end{equation}
\end{widetext}
with $b=7\zeta(3)/\pi^{3}$, the continuity equation (\ref{cont}),
and expression for the current (\ref{current}) are obtained for the
classical components of the fields (we have omitted the subscript
$cl$ for brevity). The correlators of the Gaussian noise sources may
be directly read out from the Hubbard-Stratonovich procedure and are
given by Eqs.~(\ref{xiDelta}) and (\ref{xij}).

We discuss briefly the role of the last nonlinear term in the lhs of
TDGL equation \eqref{KTDGL-NL}. On the mean-field level, i.e., being
averaged over the fluctuations of the order parameter $\langle
\Delta^{cl}_{\mathcal{K}}\rt
\Delta^{*cl}_{\mathcal{K}}(\mathbf{r}',t')\rangle\propto
L_{K}(\mathbf{r}-\mathbf{r}',t-t')$, this term leads to a
renormalization of the coefficient in front of the time derivative
(\ref{KTDGL}),
\begin{equation}
\partial_t\Delta_{\mathcal{K}} \rightarrow \left(1-c_d\,
\frac{[\xi(T)]^{4-d}}{\nu D \xi_0^2}\right)
\partial_t\Delta_{\mathcal{K}}\,,
                                        \label{viscosity-renormalization}
\end{equation}
where $\xi_0=\sqrt{D/T_c}$ and $\xi(T)=\sqrt{D/(T-T_c)}$. The
dimensionality dependent coefficient $c_{d}$ appears as the result
of the convolution between the diffuson and the Keldysh components
of the fluctuations propagator and reads as
\begin{equation}
c_{d}=\frac{7\zeta(3)}{8\pi^{3}}
\int\frac{\mathrm{d}^{d}k}{(2\pi)^{d}}
\frac{1}{(k^{2}+1)(k^{2}+1/2)},
\end{equation}
here is $c_1=0.044$, $c_2=0.012$, and $c_3=0.005$. Note that
disorder-dependent renormalization of the
$\partial_t\Delta_{\mathcal{K}}$ term  does not violate Anderson
theorem. The down renormalization of the coefficient in front of the
first time derivative is a precursor of the oscillatory
Carlson-Goldman \cite{CG} modes, appearing below $T_c$.

Finally, the TDGL equation (\ref{KTDGL}), written in the
$\mathcal{K}$-gauge may be transformed to an arbitrary gauge by the
substitution $\Delta_\mathcal{K} =\Delta e^{-2ie\mathcal{K}}$. Such
a substitution brings the vector potential $\A$ to the covariant
spatial derivative, while the the time derivative acquires the
anomalous GE term $\partial_t\to
\partial_t - 2ie\partial_t\mathcal{K}$. This way TDGL equation
(\ref{TDGL}) is obtained.

The set of equations is  simplified in the limit of the strong
Coulomb interactions.~\cite{Gorkov-Eliashberg} The latter impose the
condition of local instantaneous charge neutrality $\rho\rt=0$ and
therefore $\Div\ \J=0$. Applying this condition to the expression
for the current (\ref{current}) and neglecting for simplicity the MT
and supercurrent contributions, one finds $e^2\nu D \Div\ \mathbf{E}
= \Div\ \xi_\J$. As a result, the longitudinal component of the
electric filed $\mathbf{E}_\parallel$ is a fluctuating Gaussian
field. Employing Eqs.~(\ref{A_Kparrallel}) and (\ref{xij}), one may
translate it to the Gaussian correlator for the longitudinal
component of the vector
$\A_{\parallel\mathcal{K}}\rt=\A_{\parallel\mathcal{K}}^{ext}(t) +
\A_{\parallel\mathcal{K}}^{fluct}\rt$, where
$\A_{\parallel\mathcal{K}}^{ext}(t)$ is an externally applied
divergenceless field and the fluctuation component has the
correlator
\begin{equation}
\left\langle \A_{\parallel\mathcal{K}}^{fluct}(\mathbf{-q},-\omega)
\A_{\parallel\mathcal{K}}^{fluct}(\mathbf{q}, \omega) \right\rangle
= \frac{2T}{e^2\nu D}\, \frac{1}{(D\mathbf{q}^2)^2 +\omega^2}\, .
                                            \label{AKparallel-fluct}
\end{equation}
Exactly the same expression may be, of course, directly read out
from the normal action (\ref{S-N-alternative}) (the Maxwell part of
the action is absent in the  the strong Coulomb limit). To this end,
one needs to perform the Gaussian integration over the
$\A_{\mathcal{K}}^{q}$ component leading to the quadratic action for
the $\A_{\mathcal{K}}^{cl}$ with the correlator given by
Eq.~(\ref{AKparallel-fluct}). It is this fluctuating vector
potential which is responsible for dephasing of the Cooperon
propagators.~\cite{AAK}

%-----------------------------------------------------------------------------
%-----------------------------------------------------------------------------
\subsection{Effective action versus diagrammatic technique}
\label{Action-vs-Diagrams}

Given that there is an existing microscopic formalism for the
Aslamazov-Larkin, Maki-Thompson, and density of states diagrams, it
is important that the results from the Keldysh effective action,
formulated in the previous sections, be compared with well
established results for the corrections to the conductivity. This
comparison is the necessary check for the validity of our approach.

We start from the density of states contribution to the
conductivity. For that purpose, one uses Eq.\eqref{deltaD-DOS} and
writes conductivity correction in the form
\begin{equation}\label{sigma-DOS-start}
\delta\sigma_{DOS}=e^{2}D\left\langle\delta
\nu^{DOS}_{\mathbf{r},t}\right\rangle_{\Delta}=-\frac{7\zeta(3)e^{2}\nu
D}{4\pi^{2}T^{2}_{c}}
\left\langle|\Delta^{cl}_{\mathcal{K}}\rt|\right\rangle_{\Delta}.
\end{equation}
The averaging over the order parameter fluctuations is done easily
in the momentum space
$\left\langle|\Delta^{cl}_{\mathcal{K}}\rt|\right\rangle_{\Delta}=
\frac{8T_{c}}{\pi\nu}\sum_{\mathbf{q}\omega}L_{K}(\mathbf{q},\omega)$.
Using then explicit form of the Keldysh component for the
fluctuations propagator \eqref{L-matrix}, one finds
\begin{equation}\label{sigma-DOS-1}
\delta\sigma_{DOS}=-\frac{28\zeta(3)}{\pi^{4}}\
e^{2}D\sum_{\mathbf{q}} \int\limits^{+\infty}_{-\infty}
\mathrm{d}\omega\frac{1}{(D\mathbf{q}^{2}+\tau^{-1}_{GL})+\omega^{2}}.
\end{equation}
Further analysis of the formula \eqref{sigma-DOS-1} depends
essentially on the system effective dimensionality. As an example,
let us concentrate on the quasi-two-dimensional geometry -- metal
film with the thickness $d$, which is much smaller than the
superconductive coherence length $d\ll\xi(T)$. In this case,
momentum sum can be written as the integral according to the
substitution $\sum_{\mathbf{q}}\rightarrow\frac{1}{d}
\int\frac{\mathrm{d}\mathbf{q}^{2}}{4\pi}$. Performing remaining
integrations, with the logarithmic accuracy one finds
\begin{equation}\label{sigma-DOS-fin}
\delta\sigma_{DOS}=-\frac{7\zeta(3)e^{2}}{\pi^{4}d}
\ln\left(\frac{T_{c}}{T-T_{c}}\right).
\end{equation}
Deriving Eq.\eqref{sigma-DOS-fin}, the momentum integration was cut
at the upper limit $D\mathbf{q}^{2}_{\mathrm{max}}\sim T_{c}$.
Recall that effective action was derived under the constraint
$D\mathbf{q}^{2}\sim\omega\ll T_{c}$, thus such a regularization is
self-consistent.

We proceed with the Maki-Thompson correction to the conductivity. In
this case, one starts from the formula
$\delta\sigma_{MT}=e^{2}\nu\left\langle\delta
D^{MT}_{\mathbf{r},t}\right\rangle_{\Delta}$, uses explicit form of
the $\delta D^{MT}_{\mathbf{r},t}$ given by Eq.~\eqref{deltaD}, and
rewrites average over $\Delta$ in the momentum space. This way one
obtains
\begin{equation}\label{sigma-MT-start}
\delta\sigma_{MT}=\frac{4}{\pi}\
e^{2}D\sum_{\mathbf{q}}\int\mathrm{d}\omega\frac{T_{c}}{D\mathbf{q}^{2}
[(D\mathbf{q}^{2}+\tau^{-1}_{GL})^{2}+\omega^{2}]}.
\end{equation}
Again in the case of the two dimensional geometry, after momentum
and frequency integrations, Eq,~\eqref{sigma-MT-start} reduces to
\begin{equation}\label{sigma-MT-fin}
\delta\sigma_{MT}=\frac{e^{2}}{8d}\left(\frac{T_{c}}{T-T_{c}}\right)
\ln\left(\frac{\tau_{\phi}}{\tau_{GL}}\right),
\end{equation}
where infrared divergency in the momentum integration was cut here
\textit{by hand} introducing dephasing time
$D\mathbf{q}^{2}_{\mathrm{min}}\sim\tau^{-1}_{\phi}$. This spurious
divergency is very well known feature of the Maki-Thompson diagram.
It was regularized by Thompson introducing magnetic impurities, and
in that case dephasing time $\tau_{\phi}$ is nothing else but the
spin flip time $\tau_{s}$.

Finally, we summarize with few additional remarks. Equations
\eqref{sigma-DOS-1} and \eqref{sigma-MT-start} can be recovered from
the traditional Matsubara diagrammatic techniques after one expands
all fluctuation propagators at small frequencies and momenta,
integrates fast fermionic energies, and keeps only contribution from
zero Matsubara frequency [in our language, latter condition strictly
speaking corresponds to the long time approximation for the
distribution function Eq.\eqref{Ftime}]. To this extent, effective
action approach contains the most divergent temperature part of the
conductivity corrections, thus allows to reproduce known results.

%----------------------------------------------------------------------------------
%----------------------------------------------------------------------------------

\section{Discussion}
\label{Section-Conclusions}

We have presented a systematic way to integrate out the fermionic
degrees of freedom in a fluctuating superconductor. The underlying
assumptions for this procedure are: (i) spatial and temporal scales
of all bosonic fields are slow in comparison with $\xi_0$ and
$1/T_c$ correspondingly (but {\em not} necessarily slow in
comparison with $\xi(T)=\xi_0\sqrt{T_c/(T-T_c)}$ and $\tau_{GL}$).
(ii) The external electromagnetic fields are sufficiently small (see
Sec. \ref{section-intro} for the details), such that the fermionic
degrees of freedom are in local equilibrium. The result is the
dynamic Keldysh action written in terms of the fluctuating order
parameter and electromagnetic potentials. This action naturally
incorporates (time nonlocal) MT terms as well as anomalous GE terms,
effectively closing the discussion whether or not   TDGL theory
includes the MT effect. We have also uncovered certain nonlocal
nonlinear terms of TDGL equation (passed previously unnoticed, to
the best of our knowledge). The nonlinear coupling of the
electromagnetic fields, leading to the Altshuler-Aronov effect, may
be also directly incorporated into the scheme.

Although we did not evaluate any physical observable here, the
derived action opens the way to describe a number of phenomena. To
name a few, we mention, e.g., nonequilibrium current noise in
proximity to the critical temperature, especially the MT
contribution to noise, which would be very difficult to calculate by
any other mean. Another possible application is evaluation of the MT
dephasing time.  Extension of the theory for $T<T_c$ to describe,
e.g., the collective modes~\cite{CG} and analysis of the dynamical
regimes far from the equilibrium, are yet another fascinating
directions.

We are grateful to L.~Glazman, B.~Ivlev, M.~Khodas, I.~Lerner and
A.~Varlamov for  useful discussions. This work is supported by the
NSF Grant Nos. DMR 02-37296, DMR 04-39026, and  DMR 0405212. A.K. is
also supported by the A.~P.~Sloan foundation.

%----------------------------------------------------------------------------------
%----------------------------------------------------------------------------------

\appendix

\section{Diffuson expansion and normal action}
\label{Appendix-Diffusons}

We first focus on the part of the action \eqref{S} which is linear
in the electromagnetic potentials. There are two such terms:
$S_{\Phi}$ originating  from the trace
$\mathrm{Tr}[\check{\Phi}_{\mathcal{K}}\check{Q}_{\mathcal{K}}]$ and
$S_{\A}$ coming from the  covariant derivative
\eqref{Covariant-Derivative}. We then expand
$\check{Q}_{\mathcal{K}}$ matrix  to the linear order in deviations
from the saddle point $\check{Q}_{\mathcal{K}}=\check{\Lambda}+
\frac{1}{2}[\check{\Lambda},\check{W}]+\ldots$, and use the cyclic
property of the trace operation to obtain $S_{\Phi}=-\frac{\pi
e\nu}{4}\mathrm{Tr}\left([\check{\Phi}_{\mathcal{K}},
\check{\Lambda}]\check{W}\right)$ and  $S_{\A}=\frac{\pi e\nu
D}{4}\mathrm{Tr}\left(\check{\Lambda}\nabla\check{W}
[\check{\Xi}\check{\mathbf{A}}_{\mathcal{K}},\check{\Lambda}]\right)$
which after the integration by parts translates into
$S_{\A}=-\frac{\pi e\nu
D}{4}\mathrm{Tr}\left(\check{\Xi}\nabla\check{\mathbf{A}}_{\mathcal{K}}
- \check{\Lambda}\check{\Xi}\nabla\check{\mathbf{A}}_{\mathcal{K}}
\check{\Lambda}\right)\check{W}$. Requiring that terms linear in
variation $\check{W}$ (and linear in potentials) vanish, one arrives
at  Eq~\eqref{Vanishing-wz-condition}.

Expanding action \eqref{S} to the second order in $\check{W}$ and
keeping track of the diffusons only, one finds quadratic action of
the diffuson degrees of freedom
\begin{equation}
iS_{D}[w^{0},w^{z}]=\frac{\pi\nu}{2}\mathrm{Tr}\left\{\bar{w}^{\alpha}_{tt'}
[-D\nabla^{2}+\partial_{t}+\partial_{t'}]w^{\alpha}_{t't}\right\}\,
,
\end{equation}
where $\alpha =0,z$. This action leads to the following propagator:
\begin{equation}
\langle w^{\alpha}_{\varepsilon_{1},\varepsilon_{2}}(\mathbf{q})
\bar{w}^{\alpha}_{\varepsilon_{3},\varepsilon_{4}}(-\mathbf{q})\rangle=-\frac{2}{\pi\nu}
\frac{(2\pi)^{2}\delta_{\varepsilon_{1}-\varepsilon_{4}}
\delta_{\varepsilon_{2}-\varepsilon_{3}}}{D\mathbf{q}^{2}-
i(\varepsilon_{1}-\varepsilon_{2})}\,.
\end{equation}
%----------------------------------------------------------------------------------

Substituting the saddle point $\check{\Lambda}$  in  the
$\sigma$-model action \eqref{S}, one finds
\begin{eqnarray}\label{S-N-Intermediate}
S_{N}&=&\frac{e^{2}\nu}{2}\, \mathrm{Tr}
\left\{\check{\Phi}_{\mathcal{K}}\check{\Upsilon}
\check{\Phi}_{\mathcal{K}}\right\} \\ &-& \frac{i\pi e^{2}\nu
D}{4}\mathrm{Tr}\,
\left\{\check{\Xi}\check{\mathbf{A}}_{\mathcal{K}}
\check{\Lambda}\check{\Xi}\check{\mathbf{A}}_{\mathcal{K}}
\check{\Lambda}- \check{\Xi}\check{\mathbf{A}}_{\mathcal{K}}
\check{\Xi}\check{\mathbf{A}}_{\mathcal{K}}\right\}\,. \nonumber
\end{eqnarray}
The traces are evaluated using the following  matrix identity:
\begin{equation}\label{Matrix-Identity}
\int\limits^{+\infty}_{-\infty}\mathrm{d}\epsilon\
\mathrm{Tr}\left[\check{\Upsilon}_{\alpha}\check{\Upsilon}_{\beta}-
\check{\Upsilon}_{\alpha}
\check{\Lambda}_{\epsilon_{+}}\check{\Upsilon}_{\beta}
\check{\Lambda}_{\epsilon_{-}}\right]=8\omega
(\hat{\Pi}^{-1}_{\omega})_{\alpha\beta}\, ,
\end{equation}
where
\begin{equation}
\label{Pi-1}
\hat{\Pi}^{-1}_{\omega}=\!\left(\begin{array}{cc}0 & -1 \\
1 & 2B_{\omega}\end{array}\right)
\end{equation}
and $\check{\Upsilon}_{1}=\sigma_{0}\otimes\tau_{z}$,
$\check{\Upsilon}_{2}=\sigma_{x}\otimes\tau_{z}$, $\epsilon_\pm
=\epsilon\pm\omega/2$. The identity  is based on the   relation
between the bosonic and fermionic distribution functions:
\begin{subequations}\label{F-B-Integrals}
\begin{equation}
\int\limits^{+\infty}_{-\infty}\mathrm{d}\epsilon
(F_{\epsilon_{+}}-F_{\epsilon_{-}})=2\omega,
\end{equation}
\begin{equation}
\int\limits^{+\infty}_{-\infty}\mathrm{d}\epsilon
(1-F_{\epsilon_{+}}F_{\epsilon_{-}})=2\omega B_{\omega}.
\end{equation}
\end{subequations}
Combining  Eqs.~\eqref{S-N-Intermediate}-\eqref{F-B-Integrals}, we
arrive at normal metal action \eqref{S-N}. Finally, to find  the
macroscopic equations \eqref{cont} and \eqref{current-normal}, one
needs to express \textit{gauged} electromagnetic potentials
$\Phi_{\mathcal{K}},\mathbf{A}_{\mathcal{K}}$ in terms of
\textit{bare} ones $\Phi,\mathbf{A}$. Using
Eq.~\eqref{Kgauge-complete}, one may  find the relation between
those
\begin{subequations}\label{A-K-Phi-K}
\begin{equation}
\vec{\mathbf{A}}_{\mathcal{K}}(\mathbf{q},\omega)=
\vec{\mathbf{A}}_{\perp}(\mathbf{q},\omega)-i\mathbf{q}
\hat{\mathcal{D}}^{\,\mathbf{q}}_{\omega}
\hat{\Pi}^{-1}_{\omega}[\vec{\Phi}+
\omega(\mathbf{q}\vec{\mathbf{A}}_{\parallel})/\mathbf{q}^{2}],
\end{equation}
\begin{equation}
\vec{\Phi}_{\mathcal{K}}(\,\mathbf{q},\omega)=D\mathbf{q}^{2}
\hat{\mathcal{D}}^{\mathbf{q}}_{\omega}\sigma_{x}[\vec{\Phi}+
\omega(\mathbf{q}\vec{\mathbf{A}}_{\parallel})/\mathbf{q}^{2}],
\end{equation}
\end{subequations}
where
\begin{equation}
\hat{\mathcal{D}}^{\,\mathbf{q}}_{\omega}=\frac{1}{(D\mathbf{q}^2)^2+\omega^2}\!
\left(\!\begin{array}{cc}
2i\omega B_{\omega} & \!\! i\omega + D\mathbf{q}^{2}\\
-i\omega + D\mathbf{q}^{2}& 0
\end{array}\right).
\end{equation}

%----------------------------------------------------------------------------------
%----------------------------------------------------------------------------------
\section{Cooperon expansion and effective action}
\label{Appendix-Gradient-Expansion}

Carrying out the Cooperon  expansion, it is convenient to
distinguish several contributions into the $\sigma$-model action
\eqref{S}:
\begin{eqnarray}
S_{\sigma}[W,\vec{\Delta}_{\mathcal{K}},
\vec{\Phi}_{\mathcal{K}},\vec{\mathbf{A}}_{\mathcal{K}}]&=&
S^{a}_{\sigma}[W]+S^{b}_{\sigma}[W,\vec{\Delta}_{\mathcal{K}}] \\
&+&S^{c}_{\sigma}[W,\vec{\mathbf{A}}_{\mathcal{K}}]+
S^{d}_{\sigma}[W,\vec{\mathbf{A}}_{\mathcal{K}}]. \nonumber
\end{eqnarray}

The  $S^{a}_{\sigma}[W]$ part corresponds to the free Cooperons
which are uncoupled from both $\Delta_{\mathcal{K}}\rt$ and
$\mathbf{A}_{\mathcal{K}}\rt$. This contribution arises after one
expands trace of the gradient term
$\mathrm{Tr}(\nabla\check{Q})^{2}=
\mathrm{Tr}(\check{\Lambda}\nabla\check{W}\check{\Lambda}\nabla\check{W})=
\mathrm{Tr}(\check{\mathcal{W}}\nabla^{2}\check{\mathcal{W}})$ and
trace of the time derivative term
$\mathrm{Tr}[\check{\Xi}\partial_{t}\check{Q}]=\frac{1}{2}\mathrm{Tr}
[\check{\Xi}\partial_{t}\check{\Lambda}\check{W}^{2}]=\frac{1}{2}\mathrm{Tr}
[\check{\Xi}\partial_{t}\check{\mathcal{U}}
\check{\Lambda}_{0}\check{\mathcal{W}}\check{\mathcal{W}}\check{\mathcal{U}}]$.
Multiplying matrices, tracing them over   Keldysh-Nambu space, one
finds
\begin{equation}\label{S-a}
S^{a}_{\sigma}[W]=\frac{i\pi\nu}{4}\mathrm{Tr}\{w^{*}_{tt'}[-D\nabla^{2}-
\partial_{t}+\partial_{t'}]w_{t't}+
(w\rightarrow\bar{w})\}.
\end{equation}

The  $S^{b}_{\sigma}[W,\Delta_{\mathcal{K}}]$ part corresponds to
the coupling term between Cooperons and the order parameter. This
contribution arises from the trace
$\mathrm{Tr}[\check{\Delta}_{\mathcal{K}}\check{Q}_{\mathcal{K}}]$
after one expands $\check{Q}_{\mathcal{K}}$ to the first order in
$\check{W}$:
$\mathrm{Tr}[\check{\Delta}_{\mathcal{K}}\check{Q}_{\mathcal{K}}]=
\mathrm{Tr}[\check{\Delta}_{\mathcal{K}}\check{\Lambda}
\check{W}]=\mathrm{Tr}[\check{\mathcal{U}}\check{\Delta}_{\mathcal{K}}
\check{\mathcal{U}} \check{\Lambda}_{0}\check{\mathcal{W}}]$. After
evaluation of traces, which is done with the help of the identity
$\mathrm{Tr}_{N}[\tau_{\pm}\tau_{z}\tau_{\mp}]=\mp1$, one finds
\begin{equation}\label{S-b}
S^{b}_{\sigma}[W,\vec{\Delta}_{\mathcal{K}}]=\frac{\pi\nu}{2}\,
\mathrm{Tr} \left[ \vec{V}^{\dag}_{tt'} \vec{\mathfrak{W}}_{t't}-
\vec{\mathfrak{W}}^{\dag}_{tt'}\vec{V}_{t't}\right],
\end{equation}
where we have used vectors $\vec{\mathfrak{W}}$ and $\vec{V}$ in the
notations of Eq. \eqref{S-W}.

The c and d parts of the action are the terms which provide the
interaction vertices between the cooperons $W$ and the vector
potential $\mathbf{A}_{\mathcal{K}}$. The
$S^{c}_{\sigma}[W,\mathbf{A}_{\mathcal{K}}]$ part is linear in the
vector potential and arises from the square of the  covariant
derivative, keeping terms linear in $\mathbf{A}_{\mathcal{K}}$
contribution $$S^{c}_{\sigma}=\frac{\pi e\nu
D}{4}\mathrm{Tr}\{\nabla\check{Q}_{\mathcal{K}}
[\check{\Xi}\check{\mathbf{A}}_{\mathcal{K}},\check{Q}_{\mathcal{K}}]\}=
$$
$$\frac{\pi e\nu D}{4}\mathrm{Tr}\{\check{\Lambda}\nabla\check{W}
[\check{\Xi}\check{\mathbf{A}}_{\mathcal{K}},\check{\Lambda}\check{W}]\}=\frac{\pi
e\nu D}{4}\mathrm{Tr}\{\check{\mathcal{U}}\check{\Xi}
\check{\mathbf{A}}_{\mathcal{K}}\check{\mathcal{U}}
[\nabla\check{\mathcal{W}}, \check{\mathcal{W}}]\}.$$ After a
straightforward algebra, one finds
\begin{equation}\label{S-c}
S^{c}_{\sigma}[W,\vec{\mathbf{A}}_{\mathcal{K}}]=-\frac{\pi e\nu
D}{4}\mathrm{Tr}\{\mathcal{A}_{tt''}[w^{*}_{tt'}\nabla
w_{t't''}-\nabla w^{*}_{tt'}w_{t't''}]
\end{equation}
\[
\qquad\qquad\qquad+
\bar{\mathcal{A}}_{tt''}[\bar{w}^{*}_{tt'}\nabla\bar{w}_{t't''}-
\nabla\bar{w}^{*}_{tt'}\bar{w}_{t't''}]-c.c.\}\, ,
\]
where we have introduced notations
\begin{subequations}
\begin{equation}
\mathcal{A}_{tt'}(\mathbf{r})=
\delta_{t-t'}\mathbf{A}^{cl}_{\mathcal{K}}\rt+\frac{i}{2T}\delta'_{t'-t}
\mathbf{A}^{q}_{\mathcal{K}}\rt\ ,
\end{equation}
\begin{equation}
\bar{\mathcal{A}}_{tt'}(\mathbf{r})=\delta_{t-t'}
\mathbf{A}^{cl}_{\mathcal{K}}\rt-
\frac{i}{2T}\delta'_{t'-t}\mathbf{A}^{q}_{\mathcal{K}}\rtp\ .
\end{equation}
\end{subequations}

The quadratic in vector potential part of the action comes from
the diamagnetic term, which has the form
$S^{d}_{\sigma}[Q_{\mathcal{K}}]=-\frac{i\pi e^{2}\nu D}{4}\,
\mathrm{Tr}
\{\check{\Xi}\check{\mathbf{A}}_{\mathcal{K}}\check{Q}_{\mathcal{K}}
\check{\Xi}\check{\mathbf{A}}_{\mathcal{K}}\check{Q}_{\mathcal{K}}\}$.
One possibility is to expand one of the $\check{Q}_{\mathcal{K}}$
matrices up to the second order in $\check{W}$, while leaving the
other to be $\check{\Lambda}$, the other is to expand both of them
to the first order
\begin{equation}\label{S-d-1}
S^{d}_{\sigma}[W,\vec{\mathbf{A}}_{\mathcal{K}}]=-\frac{i\pi
e^{2}\nu D}{4}\mathrm{Tr}\, \left[ \check{\Sigma} \check{\Sigma}
\check{\mathcal{W}} \check{\mathcal{W}} + \check{\Sigma}
\check{\mathcal{W}} \check{\Sigma} \check{\mathcal{W}} \right],
\end{equation}
where
\begin{equation}
\check{\Sigma}_{\varepsilon\varepsilon'}=
\check{\mathcal{U}}_{\varepsilon}\check{\Xi}
\check{\mathbf{A}}_{\varepsilon-\varepsilon'}
\check{\mathcal{U}}_{\varepsilon'}\check{\Lambda}_{0}=
\end{equation}
\[=\left[
\begin{array}{cc} \mathbf{A}^{cl}_{\mathcal{K}}+F_{\varepsilon}
\mathbf{A}^{q}_{\mathcal{K}} &
\mathbf{A}^{q}_{\mathcal{K}}(1-F_{\varepsilon}F_{\varepsilon'})+
\mathbf{A}^{cl}_{\mathcal{K}}(F_{\varepsilon}-F_{\varepsilon'}) \\
-\mathbf{A}^{cl}_{\mathcal{K}} &
-\mathbf{A}^{cl}_{\mathcal{K}}+F_{\varepsilon'}\mathbf{A}^{q}_{\mathcal{K}}
\end{array}\right],
\]
and all vector potential have $\varepsilon-\varepsilon'$ as an
argument. Combining it with all the contributions, given by
Eqs.\eqref{S-a}-\eqref{S-c}, the full action may be conveniently
presented using matrix notations as Eq.~\eqref{S-W}.

%----------------------------------------------------------------------------------
%----------------------------------------------------------------------------------

We turn now to the derivation of the effective action,
Eq.~\eqref{S-eff}. To this end, we transform
Eq.~\eqref{w-saddle-point} into the energy-momentum representation,
then using explicit form of the $\vec{V}$ vector, given by the
formula \eqref{V}, we find for the Cooperon generators
\begin{subequations}\label{w-saddle-point-sol}
\begin{equation}
w_{\varepsilon\varepsilon'}(\mathbf{q})=-2i\frac{\Delta^{cl}_{\mathcal{K}}
(\mathbf{q},\varepsilon-\varepsilon')+
F_{\varepsilon}\Delta^{q}_{\mathcal{K}}(\mathbf{q},\varepsilon-\varepsilon')}
{D\left(\mathbf{q}-2e\mathbf{A}^{cl}_{\mathcal{K}}\right)^{2}
-i(\varepsilon+\varepsilon')},
\end{equation}
\begin{equation}
\bar{w}_{\varepsilon\varepsilon'}(\mathbf{q})=2i
\frac{\Delta^{cl}_{\mathcal{K}}(\mathbf{q},\varepsilon-\varepsilon')-
F_{\varepsilon'}\Delta^{q}_{\mathcal{K}}(\mathbf{q},\varepsilon-\varepsilon')}
{D\left(\mathbf{q}-2e\mathbf{A}^{cl}_{\mathcal{K}}\right)^{2}+
i(\varepsilon+\varepsilon')}.
\end{equation}
\end{subequations}
Note that the scale for the \textit{energy center of mass}
$\epsilon=(\varepsilon+\varepsilon')/2$ is set by the temperature
$\epsilon\sim T$. Then in most of the cases
$D\mathbf{q}^{2}\sim\omega\sim T-T_{c}$ can be ignored as compared
to  $\epsilon$ (the exception is MT term), thus one may write
instead of \eqref{w-saddle-point-sol} the approximations
\begin{subequations}
\begin{equation}
w_{\varepsilon\varepsilon'}(\mathbf{q})\approx2
\frac{\Delta^{cl}_{\mathcal{K}}(\mathbf{q},\varepsilon-\varepsilon')+
F_{\varepsilon}\Delta^{q}_{\mathcal{K}}(\mathbf{q},\varepsilon-\varepsilon')}
{\varepsilon+\varepsilon'+i0},
\end{equation}
\begin{equation}
\bar{w}_{\varepsilon\varepsilon'}(\mathbf{q})\approx2
\frac{\Delta^{cl}_{\mathcal{K}}(\mathbf{q},\varepsilon-\varepsilon')-
F_{\varepsilon'}\Delta^{q}_{\mathcal{K}}(\mathbf{q},\varepsilon-\varepsilon')}
{\varepsilon+\varepsilon'-i0},
\end{equation}
\end{subequations}
and similar equations for the conjugated fields. After the inverse
Fourier transform, one finds
\begin{subequations}\label{w-saddle-point-approx}
\begin{equation}
w_{tt'}(\mathbf{r})=-i\theta(t-t')\Delta^{cl}_{\mathcal{K}}(\mathbf{r},\tau)+
\Delta^{q}_{\mathcal{K}}(\mathbf{r},\tau)Y(t-t'),
\end{equation}
\begin{equation}
\bar{w}_{tt'}(\mathbf{r})=i\theta(t'-t)\Delta^{cl}_{\mathcal{K}}(\mathbf{r},\tau)-
\Delta^{q}_{\mathcal{K}}(\mathbf{r},\tau)Y(t-t'),
\end{equation}
\end{subequations}
where $\tau=(t+t')/2$ and
\begin{equation}
                                   \label{Y}
Y(t)=\int\limits^{+\infty}_{-\infty}\frac{\mathrm{d}\epsilon}{2\pi}
\tanh\left[\frac{\epsilon}{2T}\right]\frac{e^{-i\epsilon
t}}{\epsilon+i0}=\frac{2}{\pi}\mathrm{arctanh}\left[e^{-\pi
T|t|}\right].
\end{equation}

%----------------------------------------------------------------------------------
\subsection{$S_{GL}[\vec{\Delta}_{\mathcal{K}},
\mathbf{A}^{cl}_{\mathcal{K}}]$ part of the effective
action}\label{Appendix-S-GL}

Let us concentrate first on the part of the action which
corresponds to the diagonal blocks  ${\mathcal{C}}^{-1}$ and
$\bar{\mathcal{C}}^{-1}$. These two give identical contributions
to the action \eqref{S-W}, thus accounting for an additional
factor of $2$. We find
\begin{widetext}
\begin{equation}
S_{\sigma}[\vec{\Delta}_{\mathcal{K}}]=-2\pi i\nu\,\mathrm{Tr}\left[
\frac{\left[\Delta^{cl}_{\mathcal{K}|+}+F_{\epsilon_{+}}\Delta^{q}_{\mathcal{K}|+}\right]
\left[\Delta^{*cl}_{\mathcal{K}|-}+F_{\epsilon_{-}}\Delta^{*q}_{\mathcal{K}|-}\right]}
{D\left(\mathbf{q}-2e\mathbf{A}^{cl}_{\mathcal{K}}\right)^{2}-2i\epsilon}\right],
\end{equation}
where  we have introduced  energy integration variables as
$2\epsilon=\varepsilon+\varepsilon'$,
$\omega=\varepsilon-\varepsilon'$,
$\epsilon_{\pm}=\epsilon\pm\omega/2$, and
$\Delta_{\mathcal{K}|\pm}=\Delta_{\mathcal{K}}(\pm\mathbf{q},\pm\omega)$.
We point out here that contribution to the
$iS_{\sigma}[\vec{\Delta}_{\mathcal{K}}]$ with two classical fields
$\sim\Delta^{cl}_{\mathcal{K}}\Delta^{*cl}_{\mathcal{K}}$ is
identically zero. This is manifestation of the normalization
condition within the Keldysh formalism. Adding the term
$-2\pi\nu\,\mathrm{Tr}\left\{\Delta^{q}_{\mathcal{K}|+}
\Delta^{*q}_{\mathcal{K}|-}/[D\left(\mathbf{q}-
2e\mathbf{A}^{cl}_{\mathcal{K}}\right)^{2}-2i\epsilon]\right\}\equiv
0 $ (due to energy integration of retarded propagator), we obtain
for
$S_{\sigma}[\vec{\Delta}_{\mathcal{K}},\mathbf{A}^{cl}_{\mathcal{K}}]
+S_{\Delta}[\vec{\Delta}_{\mathcal{K}}]=
S_{GL}[\vec{\Delta}_{\mathcal{K}},\mathbf{A}^{cl}_{\mathcal{K}}]$
[cf. Eq.~(\ref{S-Delta-Phi})],
\begin{equation}
S_{GL}[\vec{\Delta}_{\mathcal{K}},
\mathbf{A}^{cl}_{\mathcal{K}}]=\nu\,\mathrm{Tr}
\left[\Delta^{*q}_{\mathcal{K}|-} L^{-1}_{R}
\Delta^{cl}_{\mathcal{K}|+}+ \Delta^{*cl}_{\mathcal{K}|-} L^{-1}_{A}
\Delta^{q}_{\mathcal{K}|+}+ \Delta^{*q}_{\mathcal{K}|-}
B(L^{-1}_{R}-L^{-1}_{A}) \Delta^{q}_{\mathcal{K}|+}\right],
\end{equation}
where we have introduced superconductive fluctuations propagator
in the form of the integral
\begin{equation}\label{L-integral}
L^{-1}_{R(A)}=-\frac{1}{\lambda}-i\int\mathrm{d}\epsilon
\frac{F_{\epsilon\mp\omega/2}}
{D\left(\mathbf{q}-2e\mathbf{A}^{cl}_{\mathcal{K}}\right)^{2}-2i\epsilon}.
\end{equation}
In what follows, we show that the latter can be reduced to the
standard form given by  equation \eqref{L}. Indeed, changing
$\varepsilon=\epsilon-\omega/2$, adding and subtracting term at zero
frequency and momentum we write
\begin{equation}\label{L-intermediate}
L^{-1}_{R}=-\frac{1}{\lambda}+\int\limits^{+\omega_{D}}_{-\omega_{D}}
\mathrm{d}\varepsilon\frac{F_{\varepsilon}}{2\varepsilon}-
i\int\limits^{+\infty}_{-\infty}\mathrm{d}\varepsilon\left[
\frac{F_{\varepsilon}}{D\left(\mathbf{q}-
2e\mathbf{A}^{cl}_{\mathcal{K}}\right)^{2}-i\omega-2i\varepsilon}+
\frac{F_{\varepsilon}}{2\varepsilon}\right],
\end{equation}
where the logarithmically divergent integral in the above formula
was cut in the standard way by the Debye frequency $\omega_{D}$.
Introducing dimensionless variable $x=\varepsilon/2T$, integrating
second term in the right hand side of Eq.~\eqref{L-intermediate} by
parts with the help of identity
$\int\limits^{\infty}_{0}\mathrm{d}x\ln(x)
\mathrm{sech}^{2}(x)=-\ln\frac{4\gamma}{\pi}$, where
$\gamma=e^{\mathbb{C}}$ with $\mathbb{C}=0.577$ being the Euler
constant, and using the definition of the metal-superconductor
transition temperature
$T_{c}=\frac{2\gamma\omega_{D}}{\pi}e^{-\frac{1}{\lambda\nu}}$, we
have for Eq.~\eqref{L-intermediate},
\begin{equation}
L^{-1}_{R}=\ln\frac{T_{c}}{T}-\frac{i}{2}
\int\limits^{+\infty}_{-\infty}\mathrm{d}x
\left[\frac{\tanh(x)}{\frac{D(\mathbf{q}-
2e\mathbf{A}^{cl}_{\mathcal{K}})^{2}
-i\omega}{4T}-ix}+\frac{\tanh(x)}{ix}\right].
\end{equation}
Using series expansion
\begin{equation}\label{tanh-series}
\tanh(x)=\sum^{\infty}_{n=0}\frac{2x}{x^{2}+x^{2}_{n}},\quad
x_{n}=\pi(n+1/2),
\end{equation}
interchanging order of summation and integration, integrating over
$x$ with the help of
\begin{equation}
\int\limits^{\infty}_{-\infty}\frac{\mathrm{d}x}{x^{2}+x^{2}_{n}}=
\frac{\pi}{x_{n}},\quad
\int\limits^{\infty}_{-\infty}\frac{x\mathrm{d}x}{\left[x^{2}+x^{2}_{n}\right]
\left[\frac{D(\mathrm{q}-2e\mathbf{A}^{cl}_{\mathcal{K}})^{2}
-i\omega}{4T}-ix\right]}=\frac{i\pi}
{\frac{D(\mathrm{q}-2e\mathbf{A}^{cl}_{\mathcal{K}})^{2}-i\omega}{4T}+x_{n}},
\end{equation}
\end{widetext}
and recalling the definition of the digamma function
\begin{equation}
\psi(x)=-C-\sum^{\infty}_{n=0}\left[\frac{1}{n+x}-
\frac{1}{n+1}\right],
\end{equation}
 one finds
\begin{equation}
L^{-1}_{R}=\ln\frac{T_{c}}{T}-\psi\left(
\frac{D(\mathbf{q}-2e\mathbf{A}^{cl}_{\mathcal{K}})^{2}-i\omega}{4\pi
T}+\frac{1}{2}\right)+ \psi\left(\frac{1}{2}\right).
\end{equation}
Finally, expanding digamma function, using $\psi'(1/2)=\pi^{2}/2$,
and transforming back to the real space and time representation
$D(\mathbf{q}-2e\mathbf{A}^{cl}_{\mathcal{K}})^{2}-i\omega\rightarrow
-D(\nabla-2ie\mathbf{A}^{cl}_{\mathcal{K}})^{2}+\partial_{t}$, one
derives Keldysh version of the Ginzburg-Landau action in the form
\eqref{S-GL} with the fluctuations propagator given by
Eq.~\eqref{L}.

%----------------------------------------------------------------------------------
\subsection{$S_{SC}[\Delta^{cl}_{\mathcal{K}},
\vec{\mathbf{A}}_{\mathcal{K}}]$ and
$S_{MT}[\Delta^{cl}_{\mathcal{K}}, \vec{\mathbf{A}}_{\mathcal{K}}]$
parts of the effective action}\label{Appendix-SC-MT}

Now we concentrate on the contributions to the effective action
\eqref{S-eff}, coming from the $\EuScript{N}$ terms of the matrix
\eqref{C}, proportional to the quantum component of the vector
potential. These contributions translate to the  action of the
form
\begin{equation}\label{S-J-intermediate}
S_{SC}[\Delta^{cl}_{\mathcal{K}},
\vec{\mathbf{A}}_{\mathcal{K}}]=\frac{i\pi\nu}{4}\mathrm{Tr}
\left[w^{*}_{tt'}\EuScript{N}_{tt'}w_{t't}+
\bar{w}^{*}_{tt'}\EuScript{N}_{tt'}\bar{w}_{t't}\right].
\end{equation}
One uses explicit form of  $\EuScript{N}$ given by
Eq.~\eqref{C-Matrix-Elements} and makes use of approximations
\eqref{w-saddle-point-approx}. As soon as
$\EuScript{N}_{tt'}\propto\delta_{t-t'}$, one may integrate over
$t'$ using regularization $\int\mathrm{d}t'
\delta(t-t')\theta(t'-t)=1/2$.  Note that deriving Eq.~\eqref{S-J}
from Eq.~\eqref{S-J-intermediate}, we kept only \textit{classical}
components of the field $\Delta$ in the Cooperons
\eqref{w-saddle-point-approx}. The \textit{quantum} components
generate interaction vertices like
$\mathrm{tr}[\mathbf{A}^{q}_{\mathcal{K}}
\Delta^{q}_{\mathcal{K}}\nabla\Delta^{*cl}_{\mathcal{K}}]$, having
more then one quantum field, are smaller than Eq.~\eqref{S-J} by the
parameter $1/T\tau_{GL}\ll1$. Indeed, one sees from
Eq.~\eqref{w-saddle-point-sol} that $\Delta^{q}_{\mathcal{K}}$ comes
in the combination with the fermionic distribution function $F$,
which according to the approximation \eqref{Ftime} brings additional
smallness by one extra power of temperature in the denominator,
which is in contrast with the term having
$\Delta^{cl}_{\mathcal{K}}$.

In the similar fashion, one derives $S_{MT}$ part of the action. We
start from
\begin{equation}\label{S-MT-Intermediate}
S_{MT}[\Delta^{cl}_{\mathcal{K}},
\vec{\mathbf{A}}_{\mathcal{K}}]=-\frac{\pi\nu}{4}
\mathrm{Tr}\left[w^{*}_{tt'}\EuScript{M}_{tt'}\bar{w}_{t't}+
\bar{w}^{*}_{tt'}\EuScript{M}_{tt'}w_{t't}\right].
\end{equation}
At this point, we again make use of approximation
\eqref{w-saddle-point-approx}. Observe that in contrast to
Eq.~\eqref{S-J-intermediate}, where we had product of either two
retarded or two advanced Cooperon fields, which restricted
integration over one of the time variables, in the case of MT
contribution \eqref{S-MT-Intermediate}, we end up with the product
between one retarded and one advanced Cooperon and the time
integration running  over the entire range $t>t'$. Precisely, this
difference between Eq.~\eqref{S-J-intermediate} and
\eqref{S-MT-Intermediate} makes contribution $S_{SC}$ to be local,
while $S_{MT}$ nonlocal. Finally, in each of the Cooperon fields
$w$, Eq.~(\ref{w-saddle-point}), one keeps only contribution with
the classical component of the order parameter. The quantum
component is again smaller by the factor of $1/T\tau_{GL}\ll1$.

%----------------------------------------------------------------------------------
%----------------------------------------------------------------------------------
\subsection{Density of states contributions to the Ginzburg-Landau action
$S_{DOS}[\Delta^{cl}_{\mathcal{K}},
\vec{\mathbf{A}}_{\mathcal{K}}]$}\label{Appendix-Dos-contribution}

There are two ways subleading DOS contributions appear in the
effective Ginzburg-Landau action. The first one, not written
explicitly in Eq.~\eqref{S-W}, is
\begin{widetext}
\begin{subequations}\label{S-W-DOS}
\begin{equation}
S^{DOS}_{\sigma}[\check
W,\vec{\Delta}_\mathcal{K},\vec{\Phi}_{\mathcal{K}},
\vec{\mathbf{A}}_{\mathcal{K}}]=\frac{i\pi\nu}{4}\mathrm{Tr}\left[
\vec{\mathfrak{W}}^{\dag}_{tt'}\left[
\begin{array}{cc}\delta\EuScript{N}^{DOS}_{tt't''} & 0 \\
0 & \delta\EuScript{N}^{DOS}_{tt't''}
\end{array}
\right] \vec{\mathfrak{W}}_{t't''}\right] ,
\end{equation}
\begin{equation}
\delta\EuScript{N}^{DOS}_{tt't''}=2e^{2}D
\left[\mathbf{A}^{q}_{\mathcal{K}}\rt
[\mathbf{A}^{cl}_{\mathcal{K}}\rt-
\mathbf{A}^{cl}_{\mathcal{K}}\rtpp]F_{t-t''}+\int\mathrm{d}t'''
\mathbf{A}^{q}_{\mathcal{K}}\rt
F_{t-t''}\mathbf{A}^{q}_{\mathcal{K}}
(\mathbf{r},t''')F_{t'''-t''}\right],
\end{equation}
\end{subequations}
Note that in order to reproduce correctly DOS contributions one
cannot use the approximate form of the fermionic distribution
function. In what follows, we deal with the part of the action
\eqref{S-W-DOS} having one classical and one quantum components of
the vector potential. The other one, having two quantum fields can
be restored using FDT. To this end, we substitute  Cooperon
generators in the form \eqref{w-saddle-point-approx} into the action
\eqref{S-W-DOS}. We keep only classical components of $\Delta$ (the
quantum one produce insignificant contributions) and account for an
additional factor of 2 due to identical contributions from $w$ and
$\bar{w}$ Cooperons.  Changing time integration variables
$t-t''=\tau$ and $t+t''=2\eta$, one finds
\begin{eqnarray}
S^{DOS}_{\sigma}=i\pi e^{2}\nu D\ \mathrm{Tr} \left[
\mathbf{A}^{q}_{\mathcal{K}}(\mathbf{r},\eta+\tau/2)
[\mathbf{A}^{cl}_{\mathcal{K}}(\mathbf{r},\eta+\tau/2)-
\mathbf{A}^{cl}_{\mathcal{K}}(\mathbf{r},\eta-\tau/2)]
F_{\tau}\times\phantom{\Delta^{cl}_{\mathcal{K}}
\left(\mathbf{r},\frac{\eta-\tau/2-t'}{2}\right)}\right.
\\ \left.\quad\quad\quad\quad
\quad\quad\times \theta(\eta+\tau/2-t')\theta(t'-\eta+\tau/2)
\Delta^{*cl}_{\mathcal{K}}
\left(\mathbf{r},\frac{\eta+\tau/2-t'}{2}\right)
\Delta^{cl}_{\mathcal{K}}
\left(\mathbf{r},\frac{\eta-\tau/2-t'}{2}\right)\right] \nonumber.
\end{eqnarray}
Note that due to the step functions, integration over $t'$ is
restricted to be in the range $\eta+\tau/2>t'>\eta-\tau/2$. Since
$F_{\tau}$ is a rapidly decreasing function of its argument, the
main contribution to the $\tau$ integral comes from the range
$\tau\sim 1/T\ll\eta$. Keeping this in mind, one makes use of the
following approximations:
$\mathbf{A}^{q}_{\mathcal{K}}(\mathbf{r},\eta+\tau/2)
[\mathbf{A}^{cl}_{\mathcal{K}}(\mathbf{r},\eta+\tau/2)-
\mathbf{A}^{cl}_{\mathcal{K}}(\mathbf{r},\eta-\tau/2)]\approx\tau
\mathbf{A}^{q}_{\mathcal{K}}(\mathbf{r},\eta)\partial_{\eta}
\mathbf{A}^{cl}_{\mathcal{K}}(\mathbf{r},\eta)$ and
$\Delta^{*cl}_{\mathcal{K}}
\left(\mathbf{r},\frac{\eta+\tau/2-t'}{2}\right)
\Delta^{cl}_{\mathcal{K}}
\left(\mathbf{r},\frac{\eta-\tau/2-t'}{2}\right)
\approx|\Delta^{cl}_{\mathcal{K}}(\mathbf{r},\eta)|^{2}$, which
allows to integrate over $t'$ explicitly
$\int\mathrm{d}t'\theta(\eta+\tau/2-t')\theta(t'-\eta+\tau/2)=\tau\theta(\tau)$.
Using fermionic distribution function \eqref{Ftime} and collecting
all factors, we find
\begin{equation}
S^{DOS}_{\sigma}=\pi e^{2}\nu DT\, \mathrm{Tr}\left[
\mathbf{A}^{q}_{\mathcal{K}}\rt
\partial_{t}\mathbf{A}^{q}_{\mathcal{K}}\rt
|\Delta^{cl}_{\mathcal{K}}\rt|^{2}\right]
\int\limits^{\infty}_{0} \frac{\tau^{2}\mathrm{d}\tau}{\sinh(\pi
T\tau)}
\end{equation}
where we set $\eta\rightarrow t$. Performing remaining integration
over $\tau$ and restoring
$S_{DOS}\sim\mathbf{A}^{q}_{\mathcal{K}}\mathbf{A}^{q}_{\mathcal{K}}$
via FDT, we arrive at
\begin{equation}\label{S-DOS}
S^{DOS}_{\sigma}= e^{2}\,\mathrm{Tr}
\left\{\delta\nu^{DOS}_{\mathbf{r},t}[\vec{\mathbf{A}}^{\dag}_{\mathcal{K}}
(\mathbf{r},t) \hat{\EuScript{T}}_{D}
\vec{\mathbf{A}}_{\mathcal{K}}(\mathbf{r},t)] \right\},\quad
\delta\nu^{DOS}_{\mathbf{r},t}=-\frac{7\zeta(3)\nu}{4\pi^{2}T^{2}}
|\Delta^{cl}_{\mathcal{K}}\rt|^{2},
\end{equation}
\end{widetext}
with $\hat{\EuScript{T}}_{D}$ given Eq.\eqref{TD}. The other source
of the DOS contributions is the matrix element $\EuScript{N}$
itself, where one has to restore fermionic distribution function,
relaxing on the approximation \eqref{Ftime}. Then in the term
$\propto\mathrm{Tr}[w_{tt'}\EuScript{N}_{tt'}w^{*}_{tt'}]$, after
one uses Eq.~\eqref{w-saddle-point-sol}, we need to keep momentum
$D\mathbf{q}^{2}$ dependance of the Cooperon and expand over
$D\mathbf{q}^2/\epsilon\ll 1$. This produces subleading contribution
such as $S_{\mathrm{eff}}\propto\frac{e^{2}\nu D}{T^{2}}
\mathrm{Tr}\{\mathbf{A}^{q}_{\mathcal{K}}D\nabla^{2}
\mathbf{A}^{cl}_{\mathcal{K}}|\Delta^{cl}_{\mathcal{K}}|^{2}\}$. As
a result, the effective action accounting for the density of states
suppression may be cast exactly into the form of
Eq.~\eqref{S-N-alternative}, where one makes the substitution
$\nu\rightarrow\nu+\delta\nu^{DOS}_{\mathbf{r},t}$.

%----------------------------------------------------------------------------------
%----------------------------------------------------------------------------------

\section{Nonlinear action
$S_{NL}[\vec{\Delta}_{\mathcal{K}}]$}\label{Appendix-S-NL}

In this section, we show how one proceed from Eq.~\eqref{S-NL-trace}
to  Eq.~\eqref{S-NL-conventional}. As was pointed out above, one
needs to keep only contributions having one quantum component of the
order parameter field. Overall, there are three possibilities to do
that in each of the Cooperon sectors $w$ and $\bar{w}$. Moreover, it
turns out that contributions coming from the $w$ and $\bar{w}$ are
identical, thus accounting for the factor of 6. We thus obtain
\begin{equation}
S_{NL}=\frac{\pi\nu}{2}\, \mathrm{Tr} [\Delta^{*cl}_{\mathcal{K}}\rt
w_{tt'}(\mathbf{r})w^{*}_{t't''}(\mathbf{r})w_{t''t}(\mathbf{r})+c.c.].
\end{equation}
We next substitute the approximate form of the Cooperon generators,
Eq.~\eqref{w-saddle-point-approx},  into this formula. In the case
of $w^{*}$, we keep quantum component of the order parameter and in
the other $w$ the classical ones,
\begin{widetext}
\begin{equation}
S_{NL}=-\frac{\pi\nu}{2}\,
\mathrm{Tr}\left[\theta(t-t')\theta(t''-t)Y(t'-t'')
\Delta^{*cl}_{\mathcal{K}}\rt
\Delta^{cl}_{\mathcal{K}}\left(\mathbf{r},\frac{t+t'}{2}\right)
\Delta^{*q}_{\mathcal{K}}\left(\mathbf{r},\frac{t'+t''}{2}\right)
\Delta^{cl}_{\mathcal{K}}\left(\mathbf{r},\frac{t+t''}{2}\right) +
c.c. \right].
\end{equation}
\end{widetext}
We change now integration variables as $t''-t=\tau$ and
$t+t''=2\eta$ and observe that the integration over $t$ is
restricted to be in the range $\eta+\tau/2>t>\eta-\tau/2$. Recall
that according to the definition \eqref{Y}, the function $Y(\tau)$
is rapidly falling on the scale $\tau\sim1/T\ll\eta$. Thus, the
major contribution to the above trace comes from the small $\tau$.
Thus, everywhere except the theta functions, one may set $t\approx
t'\approx t''=\eta$ and integrate over $t$ explicitly getting
$\tau\theta(\tau)$. Finally, using the integral
\begin{equation}
\int\limits^{\infty}_{0}\mathrm{d}\tau\ \tau\
\mathrm{arctanh}\left[e^{-\pi
T\tau}\right]=\frac{7\zeta(3)}{8\pi^{2}T^{2}},
\end{equation}
and collecting all factors, one recovers
Eq.~\eqref{S-NL-conventional}.

%----------------------------------------------------------------------------------
%----------------------------------------------------------------------------------

\end{document}